\documentclass[a4paper,11pt]{article}

\usepackage{geometry}										
\usepackage[doublespacing]{setspace}

\usepackage[english]{babel}									
\usepackage[utf8]{inputenc}									
\usepackage[T1]{fontenc}									

\usepackage{lmodern}										

\usepackage{amsmath}										
\usepackage{amstext}										
\usepackage{amsfonts}										
\usepackage{amssymb}										

\usepackage{bm}											
\usepackage{dsfont}										
\usepackage{units}										

\usepackage{textcomp}										
\usepackage{hyphenat}                                       

\usepackage{graphicx}										
\usepackage[svgnames]{xcolor}									
\usepackage{booktabs}										
\usepackage{multirow}										
\usepackage[inline]{enumitem}
\usepackage{algorithm}
\usepackage{algorithmic}

\usepackage{appendix}

\graphicspath{{./fig/}}

\usepackage{amsthm}

\theoremstyle{definition}

\usepackage{mathtools}

\newcommand{\mathbbm}[1]{\text{\usefont{U}{bbm}{m}{n}#1}} 
\newcommand{\Esp}[1]{{\mathbb E}\left[ #1 \right]}
\newcommand{\Espe}[2]{{\mathbb E}_{#1}\left[#2\right]}
\newcommand{\Var}[1]{{\rm Var}\left[ #1 \right]}
\newcommand{\Vare}[2]{{\rm Var}_{#1}\left[#2\right]}

\newcommand{\ve}[1]{\boldsymbol{#1}}
\newcommand{\acc}[1]{\left\{#1\right\}}
\newcommand{\eqdef}{\stackrel{\text{def}}{=}}
\newcommand{\con}{\,\middle|\,}
\DeclarePairedDelimiter\abs{\lvert}{\rvert}
\DeclarePairedDelimiter\Norm{\lVert}{\rVert}
\makeatletter
\let\oldabs\abs
\def\abs{\@ifstar{\oldabs}{\oldabs*}}
\let\oldnorm\Norm
\def\Norm{\@ifstar{\oldnorm}{\oldnorm*}}
\makeatother

\newcommand{\caa}{{\mathcal A}}

\newcommand{\cd}{{\mathcal D}}

\newcommand{\ch}{{\mathcal H}}
\newcommand{\cl}{{\mathcal L}}
\newcommand{\cm}{{\mathcal M}}
\newcommand{\cn}{{\mathcal N}}

\newcommand{\cu}{{\mathcal U}}
\newcommand{\cx}{{\mathcal X}}

\newcommand{\Rr}{{\mathbb R}}
\newcommand{\Nn}{{\mathbb N}}

\newcommand{\Exp}{{\rm Exp}}

\DeclareMathOperator{\LAR}{Hybrid-LAR}

\DeclareMathOperator{\OLS}{OLS}

\newcommand{\D}{\mathrm{d}}

\newcommand{\iu}{{\boldsymbol{\mathsf{u}}}}

\newlength{\HYDROsubWidth}	
\newlength{\HYDROfigHeight}	
\newlength{\HYDROmapHeight}	
\newlength{\HYDROfigHeightNew}
\usepackage{authblk}										

\title{Stochastic polynomial chaos expansions to emulate stochastic simulators}
\author[1]{Xujia Zhu\thanks{zhu@ibk.baug.ethz.ch}}
\author[1]{Bruno Sudret\thanks{sudret@ethz.ch}}
\affil[1]{Chair of Risk, Safety and Uncertainty Quantification, ETH Z\"{u}rich, Stefano-Franscini-Platz 5, 8093 Z\"{u}rich, Switzerland}

\date{\today}

\usepackage{microtype}										

\usepackage{indentfirst}									

\usepackage{caption}
\usepackage{subcaption}

\usepackage[numbers,sort&compress]{natbib}							
\usepackage{hyperref}
\hypersetup{
pdftitle = {Stochastic polynomial chaos expansions to emulate stochastic simulators},
pdfauthor = {Xujia Zhu, Bruno Sudret},
pdfsubject = {Manuscript submitted to Int. J. Uncertain. Quantificat.},
pdfkeywords = {Stochastic simulators, Surrogate modeling, Polynomial chaos expansions},
colorlinks = false,										
}

\usepackage[capitalize]{cleveref}								
\creflabelformat{equation}{#2(#1)#3}								
\crefrangelabelformat{equation}{#3(#1)#4 to #5(#2)#6}						
\labelcrefformat{subequation}{#2(#1)#3}								
\labelcrefrangeformat{subequation}{#3(#1)#4 to #5(#2)#6}					

\begin{document}

\maketitle

\begin{abstract}
In the context of uncertainty quantification, computational models are required to be repeatedly evaluated. This task is intractable for costly numerical models. Such a problem turns out to be even more severe for stochastic simulators, the output of which is a random variable for a given set of input parameters. To alleviate the computational burden, surrogate models are usually constructed and evaluated instead. However, due to the random nature of the model response, classical surrogate models cannot be applied directly to the emulation of stochastic simulators. To efficiently represent the probability distribution of the model output for any given input values, we develop a new stochastic surrogate model called \emph{stochastic polynomial chaos expansions}. To this aim, we introduce a latent variable and an additional noise variable, on top of the well-defined input variables, to reproduce the stochasticity. As a result, for a given set of input parameters, the model output is given by a function of the latent variable with an additive noise, thus a random variable. As the latent variable is purely artificial and does not have physical meanings, conventional methods (pseudo-spectral projections, collocation, regression, etc.) cannot be used to build such a model. In this paper, we propose an adaptive algorithm which does not require repeated runs of the simulator for the same input parameters. The performance of the proposed method is compared with the generalized lambda model and a state-of-the-art kernel estimator on two case studies in mathematical finance and epidemiology and on an analytical example whose response distribution is bimodal. The results show that the proposed method is able to accurately represent general response distributions, i.e., not only normal or unimodal ones. In terms of accuracy, it generally outperforms both the generalized lambda model and the kernel density estimator. 
\end{abstract}

\section{Introduction}
In modern engineering, computational models, a.k.a. simulators, are commonly 
used to simulate different operational scenarios of complex systems \emph{in 
	silico}. These models help engineers assess the reliability, control the risk, and optimize the system components in the design phase. Conventional simulators are usually deterministic: a given set of input parameters has a unique corresponding model response. In other words, repeated model evaluations with the same input values will always give identical results. In contrast, stochastic simulators return different outcomes of the model response when run twice with the same input parameters.

Stochastic simulators are widely used in engineering and applied science. The 
intrinsic stochasticity typically represents some uncontrollable effect in the 
system \cite{McNeil2005,Britton2010}. For example, in mathematical finance, Brownian motions are commonly introduced to represent stochastic effects and volatility of the stock market \cite{McNeil2005}. In epidemic simulations, additional random variables on top of the well-defined characteristic values of the population are used to simulate the stochastic spread of a disease \cite{Britton2010}.

Mathematically, a stochastic simulator can be viewed as a function
\begin{equation}\label{eq:defsto}
	\begin{split}
		\cm_s: \cd_{\ve{X}} \times \Omega &\rightarrow \Rr \\
		(\ve{x},\omega) &\mapsto \cm_s(\ve{x},\omega) ,
	\end{split}
\end{equation}
where $\cd_{\ve{X}}$ is the domain of the input parameters, and $\Omega$ denotes the probability space that represents the internal stochasticity. The latter is due to some latent random variables $\ve{\Xi}(\omega)$ which are not explicitly considered as a part of the input variables. The stochastic simulator can then be considered as a deterministic function of the input vector $\ve{x}$ and the latent variables $\ve{\Xi}$. However, it is assumed that one can only control $\ve{x}$ but not $\ve{\Xi}$ when evaluating the model. Hence, when the value of $\ve{x}$ is fixed but $\ve{\Xi}$ is generated randomly following the underlying probability distribution, the output remains random. 

In practice, each model evaluation for a fixed vector of input parameters $\ve{x}_0$ uses a particular realization of the latent variables, i.e., a particular $\omega_0 \in \Omega$ that is usually controlled by the random seed. Thus, it provides only one realization of the output random variable. In order to fully characterize the associated distribution of $\cm_s(\ve{x}_0,\cdot)$, it is necessary to repeatedly run the stochastic simulator with the same input parameters $\ve{x}_0$. The various output values obtained by this procedure are called \emph{replications} in the sequel.

In the context of uncertainty quantification or optimization, various input values should be investigated. To this aim, multiple runs of the simulator are needed for many different inputs and for many replications. This becomes impracticable for high-fidelity costly numerical models. In this context, surrogate models have received tremendous attention in the past two decades. A surrogate model is a proxy of the original model constructed from a limited number of model runs. However, standard surrogate models such as polynomial chaos expansions \cite{Ghanembook2003} and Gaussian processes \cite{Rasmussen2006} that have been successfully developed for deterministic simulators are not directly applicable to emulating stochastic simulators due to the random nature of the latter.

In the past decade, large efforts have been dedicated to estimating some summary quantities of the response distribution which are deterministic functions of the input. 

For the mean and variance of the response distribution, Ankenman et al. \cite{Ankenman2010} proposed using replications to estimate the mean and variance for various input values. The mean function is represented by a Gaussian process, for which the variance estimated from the replications is cast as a heteroskedastic effect. Marrel et al. \cite{Marrel2012} modeled both the mean and variance by Gaussian processes. The estimation procedure is similar to the feasible generalized least-squares \cite{Wooldridge2013} that consists in alternating between fitting the mean from the data and the variance from the residuals. This approach does not require replications. Binois et al. \cite{Binois2018} proposed jointly optimizing the likelihood to represent the mean and variance by Gaussian processes, which is mainly designed for data with replications. 

To estimate the quantiles of the response distribution, Koenker and Bassett \cite{Koenker1978} proposed optimizing the \emph{check function}, which established the quantile regression method. Plumlee and Tuo \cite{Plumlee2014} suggested estimating the quantiles by performing replications and building a Gaussian process from the estimated quantiles. The reader is referred to Torossian et al. \cite{Torossian2020} for a detailed review.

The methods listed above produce only targeted summary quantities. However, far less literature has been devoted to the emulation of the entire probability distribution function of the response random variable for a given input. Three types of methods can be found in the literature. 

Moutoussamy et al. \cite{Moutoussamy2015} proposed using replications to characterize the response distribution for different input values. Then, the fitted distributions (based on replications) for the discrete input values can be extended to the entire input space by parametric or nonparametric techniques. Since this approach capitalizes on replications for local inference, it is necessary to generate many replications to obtain an accurate surrogate \cite{ZhuIJUQ2020}, i.e., in the order of $10^{3}-10^{4}$ \cite{Browne2016}. 

In the second approach, a stochastic simulator is considered as a random field indexed by the input variables \cite{Azzi2019,LuethenPREM2022}. When fixing the internal stochasticity $\omega$ in \cref{eq:defsto}, the stochastic simulator is a mere deterministic function of $\ve{x}$, called \emph{a trajectory}. This function can be emulated by standard surrogate methods. Collecting different trajectories, one can approximate the underlying random field using Karhunen--Lo\`eve expansions. Therefore, it is necessary to fix the internal randomness to apply this approach, which is practically achieved by controlling the random seed. 

The third type of methods is referred to as the statistical approach and does not require replications or manipulating the random seed. If the response distribution belongs to the exponential family, generalized linear models \cite{McCullagh1989} and generalized additive models \cite{Hastie1990} can be efficiently applied. For arbitrary types of response distributions, nonparametric estimators developed in statistics can be applied, namely kernel density estimators \cite{Fan1996,Hall2004} 
and projection estimators \cite{Efromovich2010}. However, nonparametric estimators are known to suffer from the \emph{curse of dimensionality}, which indicates that the necessary amount of data increases drastically with increasing input dimensionality. To balance between very restrictive parametric assumptions and nonparametric approaches, Zhu and Sudret \cite{ZhuSIAMUQ2021,ZhuRESS2021} proposed using generalized lambda distributions to approximate the response distributions. The four distribution parameters are seen as functions of the input and further represented by polynomial chaos expansions. The main limitation of this approach is that it cannot produce multimodal distributions, however.

In this paper, we develop an original approach that directly emulates the functional representation in \cref{eq:defsto}. More precisely, we extend the classical polynomial chaos expansions to emulating stochastic simulators. We introduce a latent variable and a noise variable to reproduce the random behavior of the model output. We develop an adaptive method to construct such a surrogate model. This novel stochastic surrogate is parametric and shown to be not limited to unimodal distributions. 

The remainder of the paper is organized as follows. In \Cref{sec:PCE}, we first review the standard polynomial chaos representations. In \Cref{sec:SPCE}, we present a novel formulation named \emph{stochastic polynomial chaos expansions} which is meant for stochastic simulators. In \Cref{sec:estimation}, we present the algorithms to adaptively build such a surrogate from data without the need for replications. We illustrate the performance of the proposed method on a complex analytical example and on case studies from mathematical finance and epidemiology in \Cref{sec:examples}. Finally, we conclude the main findings of the paper and provide outlooks for future research in \Cref{sec:conclusion}.

\section{Reminder on polynomial chaos expansions}
\label{sec:PCE}
Polynomial chaos expansions (PCEs) have been widely used in the last two decades to emulate the response of deterministic simulators in many fields of applied science and engineering. Consider a deterministic model $\cm_d$ which is a function that maps the input parameters $\ve{x} = \left(x_1,x_2,\ldots,x_M\right)^T \in \cd_{\ve{X}} \subset\Rr^M$ to the scalar output $y= \cm_d(\ve{x})\in \Rr$. In the context of uncertainty 
quantification, the input vector $\ve{x}$ is affected by uncertainties 
and thus modeled by a random vector $\ve{X}$ with prescribed joint probability 
density function (PDF) denoted by $f_{\ve{X}}$. In the sequel, 
we focus on the case where the input parameters are independent for simplicity. Therefore, the joint PDF is expressed by
\begin{equation}
	f_{\ve{X}}(\ve{x}) = \prod_{j=1}^{M}f_{X_j}(x_j),
\end{equation}
where $f_{X_j}$ is the marginal PDF of the input random variable $X_j$. Note that in the case where the input vector $\ve{X}$ has dependent components, it is always possible to transform them into independent ones using the Nataf or Rosenblatt transform \cite{Nataf1962,Rosenblatt1952,BlatmanPEM2010}.

Because of the randomness in the input, the model response $Y = \cm_d(\ve{X})$ 
becomes a random variable. Provided that $Y$ has a finite variance, i.e., $\Var{Y}<+\infty$, the function $\cm_d$ belongs to the Hilbert space $\ch$ of square-integrable functions with respect to the inner product
\begin{equation}
	\langle u,v\rangle_{\ch} \eqdef \Esp{u(\ve{X})v(\ve{X})} = 
	\int_{\cd_{\ve{X}}} 
	u(\ve{x})v(\ve{x})f_{\ve{X}}(\ve{x}) \D\ve{x}.
\end{equation}
Under certain conditions on the joint PDF $f_{\ve{X}}$ \cite{Ernst2012}, 
the Hilbert space $\ch$ possesses a polynomial basis. As a result, $\cm_d$ can 
be represented by an orthogonal series expansion
\begin{equation}\label{eq:PCE}
	\cm_d(\ve{x}) = \sum_{\ve{\alpha} \in \Nn^M} c_{\ve{\alpha}} 
	\psi_{\ve{\alpha}}(\ve{x}),
\end{equation}
where $c_{\ve{\alpha}}$ is the coefficient associated with the basis function 
$\psi_{\ve{\alpha}}$ that is defined by the multi-index $\ve{\alpha}$. More 
precisely, the multivariate basis function $\psi_{\ve{\alpha}}$ is given by a tensor product of univariante polynomials
\begin{equation}\label{eq:PCE_basis}
	\psi_{\ve{\alpha}}(\ve{x}) = \prod_{j=1}^{M} \phi^{(j)}_{\alpha_j}(x_j),
\end{equation}
where $\alpha_j$ indicates the degree of $\psi_{\ve{\alpha}}(\ve{x})$ in its 
$j$-th component $x_j$, and $\acc{\phi^{(j)}_k:k\in\Nn}$ is the orthogonal polynomial basis with respect to the marginal distribution $f_{X_j}$ of $X_j$, which satisfies
\begin{equation}\label{eq:unPCE_ortho}
	\Esp{\phi^{(j)}_{k}(X_j)\,\phi^{(j)}_{l}(X_j)}=\delta_{kl}.
\end{equation}
In the equation above, the Kronecker symbol $\delta_{kl}$ is such that $\delta_{kl}=1$ if $k=l$ and $\delta_{kl}=0$ otherwise.

Following \cref{eq:PCE_basis}, the multivariate polynomial basis is defined from univariate orthogonal polynomials that depend on the corresponding marginal distribution. For uniform, normal, gamma and beta distributions, the associated orthogonal 
polynomial families are known analytically \cite{Xiu2002}. For arbitrary 
marginal distributions, such a basis can be iteratively computed by the 
\emph{Stieltjes procedure} \cite{Gautschi2004}.

The spectral representation in \cref{eq:PCE} involves an infinite sum of terms. 
In practice, the series needs to be truncated to a finite sum. The standard truncation scheme is defined by selecting all the polynomials whose total degree is small than a given value $p$, i.e., $\caa^{p,M} = \acc{\ve{\alpha}\in \Nn^M, \sum_{j=1}^{M} \alpha_j \leq p}$. However, this will provide a large number of terms for big values of $p$ and $M$. A more flexible scheme is the hyperbolic ($q$-norm) truncation scheme 
\cite{SudretJCP2011}:
\begin{equation}\label{eq:qnorm}
	\caa^{p,q,M} = \acc{\ve{\alpha}\in \Nn^M, \|\ve{\alpha}\|_{q} \leq p},
\end{equation}
where $p$ is the maximum polynomial degree, and $q \in (0,1]$ defines 
the quasi-norm $\Norm{\ve{\alpha}}_q = 
\left(\sum_{j=1}^{M}\abs{\alpha_j}^q\right)^{1/q}$. This truncation scheme allows excluding high-order interactions among the input variables but keeps 
univariate effects up to degree $p$. Note that with $q=1$, we recover the full 
basis of total degree less than $p$. 

To estimate the coefficients in \cref{eq:PCE}, one popular approach relies on minimizing the mean-squared error between the model response and the surrogate model. The basic method applies ordinary least-squares (OLS) with a given set of basis (e.g., defined by a truncation scheme) \cite{Berveiller2006}. In this approach, the model is evaluated on a number of points called the \emph{experimental design} $\cx = \acc{\ve{x}^{(1)},\ldots,\ve{x}^{(N)}}$. The associated model responses are gathered into $\ve{y} = \acc{y^{(1)},\ldots,y^{(N)}}$ with $y^{(i)} = \cm\left(\ve{x}^{(i)}\right)$. The basis functions (and thus the coefficients) can be arranged by ordering the multi-indices $\acc{\ve{\alpha}_j}_{j=1}^P$. The regression matrix $\ve{\Psi}$ is defined by $\Psi_{ij} = \psi_{\ve{\alpha}_j}\left(\ve{x}^{(i)}\right)$. By minimizing the mean-squared error between the original model and the surrogate on the experimental design, the OLS estimator is given by
\begin{equation}\label{eq:OLS}
	\hat{\ve{c}} = \arg\min_{\ve{c}}  \Norm{\ve{y} - \ve{\Psi}\ve{c}}^2_2
\end{equation}

With increasing polynomial degree or input dimension, the number of coefficients increases drastically. As a consequence, a large number of models runs are necessary to guarantee a good accuracy, which becomes intractable for costly simulators. To solve this problem, Blatmann and Sudret \cite{SudretJCP2011}, Doostan and Owhadi \cite{Doostan2011}, Babacan et al. \cite{Babacan2010} developed methods to build sparse PCEs by only selecting the most influential polynomials. The reader is referred to the review papers by L\"uthen et al. \cite{Luethen2021,LuethenIJUQ2021} for more details.

\section{Stochastic polynomial chaos expansions}
\label{sec:SPCE}
\subsection{Introduction}
Let us now come back to stochastic simulators. It would be desirable to have a spectral expansion such as \cref{eq:PCE} for stochastic simulators. Indeed, the standard PCE has numerous features such as close-to-zero-cost model evaluations, and clear interpretation of the coefficients in terms of sensitivity analysis \cite{SudretRESS2008b}. However, because the spectral expansion in \cref{eq:PCE} is a deterministic function of the input parameters, it cannot be directly used to emulate stochastic simulators. 

Considering the randomness in the input variables, the output of a stochastic simulator is a random variable. The randomness of the latter comes from both the intrinsic stochasticity and the uncertain inputs. When fixing the input parameters, the model response remains random. For the purpose of clarity, we denote by $Y_{\ve{x}}$ the random model response for the input parameters $\ve{x}$ and by $Y$ the model output containing all the uncertainties: following \cref{eq:defsto}, we have
\begin{equation}
	Y_{\ve{x}} \eqdef \cm_s(\ve{x},\omega), \quad Y \eqdef \cm_s(\ve{X}(\omega),\omega).
\end{equation}

From a probabilistic perspective, $Y_{\ve{x}}$ is equivalent to the conditional random variable $Y \mid \ve{X}=\ve{x}$. Let $F_{Y \mid \ve{X}}\left(y\con \ve{x}\right)$ denote the associated cumulative distribution function (CDF). By using the probability integral transform, we can transform \emph{any} continuous random variable $Z$ to the desired distribution, that is
\begin{equation}\label{eq:isoprob}
	Y_{\ve{x}} \stackrel{\D}{=} F^{-1}_{Y\mid \ve{X}}\left(F_{Z}\left(Z\right)\con\ve{x}\right)
\end{equation}
where $F_{Z}$ is the CDF of $Z$. The equality in \cref{eq:isoprob} is to be understood \emph{in distribution}, meaning that two random variables on the left- and right-hand side follow the same distribution. In \cref{eq:isoprob}, the right-hand side is a deterministic function of both $\ve{x}$ and $z$. As a result, assuming that $Y$ has a finite variance, we can represent this function using a PCE in the $(\ve{X},Z)$ space, that is,
\begin{equation}\label{eq:isoprobPCE}
	F^{-1}_{Y\mid \ve{X}}\left(F_{Z}\left(Z\right)\con\ve{X}\right) = \sum_{\ve{\alpha} \in \Nn^{M+1}} 
	c_{\ve{\alpha}} \psi_{\ve{\alpha}}\left(\ve{X},Z\right).
\end{equation}
For a given vector of input parameters $\ve{x}$, the expansion is a function of the artificial latent variable $Z$, thus a random variable
\begin{equation}\label{eq:latent1}
	Y_{\ve{x}} \stackrel{\D}{=} \sum_{\ve{\alpha} \in \Nn^{M+1}} 
	c_{\ve{\alpha}} \psi_{\ve{\alpha}}\left(\ve{x},Z\right).
\end{equation}
Then, we apply a truncation scheme $\caa$ (e.g., \cref{eq:qnorm}) to reduce \cref{eq:latent1} to a finite sum
\begin{equation}\label{eq:latent2}
	Y_{\ve{x}} \stackrel{\D}{\approx} \tilde{Y}_{\ve{x}} = \sum_{\ve{\alpha} \in \caa} 
	c_{\ve{\alpha}} \psi_{\ve{\alpha}}\left(\ve{x},Z\right).
\end{equation}

Even though \cref{eq:latent2} is derived from \cref{eq:isoprobPCE}, it is more general. \Cref{eq:isoprob} offers one way to represent the response distribution by a transform of a latent variable. But many other transforms can achieve the same goal. For example, using $Z \sim \cn(0,1)$, both $\mu(\ve{x})+\sigma(\ve{x}) Z$ and $\mu(\ve{x})-\sigma(\ve{x}) Z$ can represent the stochastic simulator defined by $Y_{\ve{x}} \sim \cn(\mu(\ve{x}),\sigma(\ve{x}))$. Because we are interested in the response distribution, \cref{eq:latent2} only requires that the polynomial transform of the latent variable produces a distribution that is close to the response distribution, but the transform does not need to follow \cref{eq:isoprobPCE} exactly. Note that the latent variable $Z$ is only introduced to reproduce the stochasticity, but it does not allow us to represent the detailed data generating process of the simulator though. In other words, the PCE in \cref{eq:latent2} cannot emulate the response for a particular replication, yet it provides a representation of the distribution of $Y_{\ve{x}}$.

\subsection{Potential issues with the formulation in \cref{eq:latent2}}
\label{sec:discuss}
Building a PCE by least-squares as presented in \cref{sec:PCE} requires evaluating the deterministic function to surrogate, which, in the case of stochastic simulators, is the left-hand side of \cref{eq:isoprobPCE}. However, it is practically impossible to evaluate such a function, as the response distribution $F^{-1}_{Y\mid \ve{X}}$ is unknown. One common way to fit the latent variable model defined in \cref{eq:latent2} is maximum likelihood estimation \cite{Everitt1984,Desceliers2006a}. In this section, we show some potential problems associated with a standard use of this method for building \cref{eq:latent2}, which calls for a novel fitting algorithm.

According to the definition in \cref{eq:latent2}, $\tilde{Y}_{\ve{x}}$ is a function of $Z$. Denote $f_Z(z)$ the PDF of $Z$ and $\cd_Z$ the support of $Z$. Based on a change of variable \cite{Jacod2004}, we can obtain the PDF of $\tilde{Y}_{\ve{x}}$, which is denoted by $f_{\tilde{Y}_{\ve{x}}}(y;\ve{x},\ve{c})$. As a result, the (conditional) likelihood function of the coefficients $\ve{c}$ for a data point $(\ve{x},y)$ is given by
\begin{equation}\label{eq:likelihood1}
	l(\ve{c};\ve{x},y) = f_{\tilde{Y}_{\ve{x}}}(y;\ve{x},\ve{c}).
\end{equation}

Now, let us consider an experimental design $\cx = \acc{\ve{x}^{(1)},\ldots,\ve{x}^{(N)}}$. The stochastic simulator is assumed to be evaluated \emph{once} for each point $\ve{x}^{(i)}$, yielding $\ve{y} = \acc{y^{(1)},\ldots,y^{(N)}}$ with $y^{(i)} = \cm_s\left(\ve{x}^{(i)},\omega^{(i)}\right)$. Note that here we do not control the random seed, so the model outcomes for different values of $\ve{x}$ are \emph{independent}. Thus, the likelihood function can be computed by the product of $l\left(\ve{c};\ve{x}^{(i)},y^{(i)}\right)$ over the $N$ data points. As a result, the maximum likelihood estimator is given by
\begin{equation}\label{eq:MLE1}
	\hat{\ve{c}} = \arg\max_{\ve{c}} \sum_{i=1}^{N}  \log l\left(\ve{c};\ve{x}^{(i)},y^{(i)}\right).
\end{equation}
\Cref{eq:MLE1} commonly serves as a basic approach for fitting parametric statistical models (including stochastic surrogates) \cite{McCullagh1989,Hastie2001,ZhuSIAMUQ2021}. However, the likelihood function of the latent PCE defined in \cref{eq:latent2} is unbounded and can reach $+\infty$, making the maximization problem \cref{eq:MLE1} ill-posed. 

To illustrate the issue, let us consider a simple stochastic simulator without input variables, which gives a realization of $Y$ upon each model evaluation. Hence, the surrogate in \cref{eq:latent2} contains only the latent variable $Z$, that is, $\tilde{Y} = g\left(Z\right) = \sum_{\ve{\alpha} \in \caa} c_{\ve{\alpha}} \psi_{\ve{\alpha}}\left(Z\right)$. For simplicity, let $g(z)$ be a second-degree polynomial expressed by monomials $g(z) = a_1 z^2 + a_2 z + a_3$. Note that there is a one-to-one mapping between monomials and full polynomial chaos basis, so one can map $\ve{a} = (a_1,a_2,a_3)^T$ to $\ve{c}$ through a change of basis. Using a change of variable \cite{Jacod2004}, the PDF of $\tilde{Y}$ is
\begin{equation}\label{eq:problem}
	f_{\tilde{Y}}(y) = \frac{f_{Z}(z)}{\abs{g^\prime(z)}} \mathbbm{1}_{g(z)}(y),
\end{equation} 
where $\mathbbm{1}$ is the indicator function, and $g^{\prime}$ denotes the derivative of $g$. For a given $y_0$, certain choices of $\ve{a}$ can make any given $z_0$ with $f_{Z}(z_0) \neq 0$ satisfy $g(z_0)=y_0$ and $g^\prime(z_0) = 0$:
\begin{equation}\label{eq:problemdemo}
	\begin{cases}
		g(z_0)=y_0 \\
		g^\prime(z_0) = 0
	\end{cases} \Rightarrow \;
	\begin{cases}
		a_1 z_0^2 + a_2 z_0 + a_3 - y_0=0 \\
		2 a_1 z_0 + a_2 = 0
	\end{cases} \Rightarrow \;
	\begin{cases}
		-z_0^2\,a_1^2 + a_3 - y_0 = 0 \\
		a_2 = -2z_0\,a_1
	\end{cases}.
\end{equation}
The system of equations in \cref{eq:problemdemo} is underdetermined for $\ve{a}$. Therefore, there are infinite combinations of the coefficients $\ve{a}$, and therefore of $\ve{c}$, such that the denominator of \cref{eq:problem} is zero and the numerator is non-zero, which gives $f_{\tilde{Y}}(y_0) = +\infty$. Consequently, the maximum likelihood estimation will always produce a certain vector $\ve{c}$ that makes the likelihood reach $+\infty$. 

As a conclusion, the surrogate ansatz of \cref{eq:latent2} can produce non-smooth conditional PDFs with singularity points where $f_{\tilde{Y}_{\ve{x}}}$ tends to infinity. Consequently, the standard maximum likelihood estimation would fail. 

\subsection{Formulation of stochastic polynomial chaos expansions}
In the previous section, we discussed some potential problems of the model defined in \cref{eq:latent2}. To regularize the optimization problem in \cref{eq:MLE1} and smooth out the produced PDFs, we introduce an additive noise variable $\epsilon$, and define the stochastic surrogate as follows:
\begin{equation}\label{eq:SPCE}
	Y_{\ve{x}} \stackrel{\rm d}{\approx} \tilde{Y}_{\ve{x}} = \sum_{\ve{\alpha} \in \caa} c_{\ve{\alpha}} 
	\psi_{\ve{\alpha}}\left(\ve{x},Z\right) + \epsilon,
\end{equation}
where $\epsilon$ is a centered Gaussian random variable with standard deviation $\sigma$, i.e., $\epsilon \sim \cn(0,\sigma^2)$. With this new formulation, the response PDF of the stochastic surrogate is a convolution of that of the PCE and the Gaussian PDF of $\epsilon$. Let $G_{\ve{x}} = \sum_{\ve{\alpha} \in \caa} c_{\ve{\alpha}} \psi_{\ve{\alpha}}\left(\ve{x},Z\right)$. The PDF of $\tilde{Y}_{\ve{x}} = G_{\ve{x}}+\epsilon$ reads
\begin{equation}
	f_{\tilde{Y}_{\ve{x}}}(y) = (f_{G_{\ve{x}}}*f_{\epsilon})(y) =\int_{-\infty}^{+\infty} f_{G_{\ve{x}}}(y-t)f_{\epsilon}(t)\D t.
\end{equation}
Using H\"older's inequality, the above integral is bounded from above by
\begin{equation}\label{eq:holder}
	\Norm{f_{G_{\ve{x}}}}_1\,\Norm{f_{\epsilon}}_{\infty} = \Norm{f_{\epsilon}}_{\infty} = \frac{1}{\sigma \sqrt{2\pi}},
\end{equation}
meaning that the PDF of $\tilde{Y}_{\ve{x}}$ and the associated likelihood function are bounded.

To illustrate the role of the additive noise variable in \cref{eq:SPCE}, let us consider a random variable $Y$ with bimodal distribution to be represented by
\begin{equation}
	Y \stackrel{\rm d}{\approx} \sum_{\ve{\alpha} \in \caa} c_{\ve{\alpha}} 
	\psi_{\ve{\alpha}}\left(Z\right) + \epsilon,
\end{equation}
where the latent variable $Z$ follows a standard normal distribution and $\epsilon \sim \cn(0,\sigma)$. In the case $\sigma =0$ (the noise term vanishes), we build the model by applying a standard algorithm such as least-angle regression (LAR) \cite{SudretJCP2011} to the probability integral transform $F^{-1}_Y(F_Z(Z))$. When the regularization term $\epsilon$ is added, maximum likelihood estimation can be used (see \cref{sec:MLE} for details) to construct the surrogate.

\begin{figure}[!htbp]
	\centering
	\includegraphics[width=0.8\linewidth, keepaspectratio]{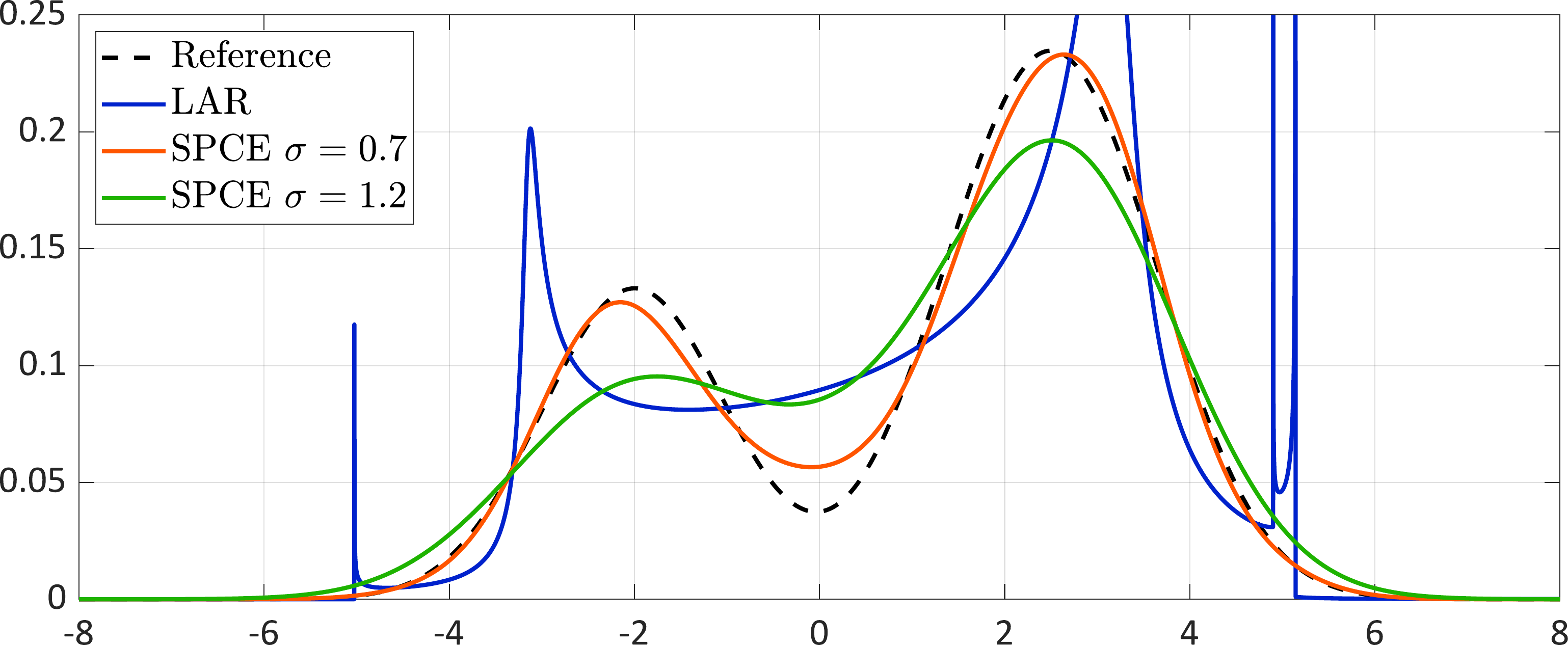}
	\caption{Emulating a bimodal distribution. The blue line corresponds to the result of using LAR to represent directly the probability integral transform (without regularization term). The red and green lines are the results of maximum likelihood estimation for two different values of $\sigma$.}
	\label{fig:unibimodal}
\end{figure}

\Cref{fig:unibimodal} shows the original (reference) PDF, and the ones obtained by LAR ($\sigma = 0$) and by the stochastic PCE for two different values of $\sigma$. It is observed that the PDF obtained by LAR has singularity points, which confirms the analysis in \cref{sec:discuss}, whereas the proposed noise term regularizes the PDFs. Moreover, LAR is applied directly to the probability integral transform which in practice is unknown. In contrast, the maximum likelihood estimation does not require knowing the values of $Z$ (in this example, only the realizations of $Y$ are used). Finally, the value of $\sigma$ affects the accuracy of the model. Hence, $\sigma$ is an additional parameter of the model that must also be fitted to the data to get the optimal approximation. The fitting procedure is detailed in the next section.

\section{Fitting the stochastic polynomial chaos expansion}
\label{sec:estimation}
To construct a stochastic PCE defined in \cref{eq:SPCE}, one needs to estimate both the coefficients $\ve{c}$ and the standard deviation $\sigma$ of the noise variable. In this section, we present a method to calibrate these parameters from data without replications. Moreover, we propose an algorithm that adaptively selects an appropriate distribution for the latent variable $Z$ and truncation scheme $\caa$.

\subsection{Maximum likelihood estimation}
\label{sec:MLE}
Let us assume for a moment that the standard deviation $\sigma$ of the noise variable is given (the estimation of $\sigma$ will be investigated separately in \Cref{sec:cv}). From \cref{eq:SPCE}, we see that our surrogate response $\tilde{Y}_{\ve{x}}$ is the sum of a polynomial function of $(\ve{x},z)$ and the noise variable $\epsilon$. Therefore, its PDF can be computed by
\begin{equation}\label{eq:PDF}
	\begin{split}
		f_{\tilde{Y}_{\ve{x}}}(y) &= \int_{\cd_{Z}} f_{\tilde{Y}_{\ve{x}} \mid Z}(y \mid z) f_Z(z)\D z\\
		&=\int_{\cd_{Z}}  \frac{1}{\sigma}
		\varphi\left(\frac{y - \sum_{\ve{\alpha} \in \caa} 
			c_{\ve{\alpha}} \psi_{\ve{\alpha}}(\ve{x},z)}{\sigma}\right) \, f_{Z}(z) \, \D z,
	\end{split}
\end{equation}
since $\tilde{Y}_{\ve{x}} \mid Z=z$ is a Gaussian random variable with mean value $\sum_{\ve{\alpha} \in \caa} 
c_{\ve{\alpha}} \psi_{\ve{\alpha}}(\ve{x},z)$ and variance $\sigma^2$ according to \cref{eq:SPCE}. In this equation, $\varphi$ stands for the standard normal PDF. Therefore, for a given data point $(\ve{x},y)$, the likelihood of the parameters $\ve{c}$ conditioned on $\sigma$ reads
\begin{equation}\label{eq:likelihood}
	l(\ve{c};\ve{x},y,\sigma) = \int_{\cd_{Z}} \frac{1}{\sqrt{2\pi}\sigma} 
	\exp\left(-\frac{\left(y - \sum_{\ve{\alpha} \in \caa} c_{\ve{\alpha}} 	
		\psi_{\ve{\alpha}}(\ve{x},z)\right)^2}{2\sigma^2}\right) f_{Z}(z) \D z.
\end{equation} 
In practice, we can use numerical integration schemes, namely Gaussian quadrature \cite{Golub1969}, to efficiently evaluate this one-dimensional integral, that is
\begin{equation}\label{eq:quadrature}
	l(\ve{c};\ve{x},y,\sigma) \approx \tilde{l}(\ve{c};\ve{x},y,\sigma)= 
	\sum_{j=1}^{N_Q} \frac{1}{\sqrt{2\pi}\sigma}\exp\left(-\frac{\left(y - 
		\sum_{\ve{\alpha} \in \caa} c_{\ve{\alpha}} 			
		\psi_{\ve{\alpha}}(\ve{x},z_j)\right)^2}{2\sigma^2}\right) w_j ,
\end{equation}
where $N_Q$ is the number of integration points, $z_j$ is the $j$-th integration point, and $w_j$ is the corresponding weight, both associated to the weight function $f_Z$. Based on \cref{eq:quadrature} and the available data $(\cx,\ve{y})$, the PCE coefficients $\ve{c}$ can be fitted using the maximum likelihood estimation (MLE)
\begin{equation}\label{eq:MLE}
	\hat{\ve{c}} = \arg\max_{\ve{c}} \; \sum_{i}^{N}
	\log\left(\tilde{l}\left(\ve{c};\ve{x}^{(i)},y^{(i)},\sigma\right)\right).
\end{equation}
The gradient of \cref{eq:quadrature}, and therefore of \cref{eq:MLE}, can be derived analytically. Hence, we opt for the derivative-based BFGS quasi-Newton method \cite{Fletcher1987} to solve this optimization problem.

\subsection{Starting point for the optimization}
The objective function to optimize in \cref{eq:MLE} is highly nonlinear. As a result, a good starting point is necessary to ensure convergence. According to the properties of the polynomial chaos basis functions, the mean function of a stochastic PCE can be expressed as
\begin{equation}\label{eq:mean}
	\tilde{m}(\ve{x}) \eqdef \Esp{\tilde{Y}_{\ve{x}}} = 
	\Espe{Z,\epsilon}{\sum_{\ve{\alpha} \in \caa} c_{\ve{\alpha}} 
		\psi_{\ve{\alpha}}\left(\ve{x},Z\right) + \epsilon} = \sum_{\ve{\alpha} \in \caa, \alpha_{z} = 0 }c_{\ve{\alpha}} \psi_{\ve{\alpha}}(\ve{x}),
\end{equation}
where $\alpha_{z}$ is the degree of the univariate polynomial in $Z$. \Cref{eq:mean} contains all the terms without $Z$, as indicated by $\alpha_{z} = 0$. We define this set of multi-indices as
\begin{equation}\label{eq:cam}
	\caa_m = \acc{\ve{\alpha} \in \caa: \alpha_{z} = 0}.
\end{equation}

Another surrogate $\hat{m}(\ve{x})$ of the mean function can be obtained by using standard (or sparse) regression to directly fit the following expansion:
\begin{equation}\label{eq:meanfit}
	m(\ve{x}) \eqdef \Esp{Y_{\ve{x}}} \approx \hat{m}(\ve{x}) \eqdef \sum_{\ve{\alpha} \in \caa_m} c^m_{\ve{\alpha}} \psi(\ve{x})
\end{equation}
The obtained coefficients $\ve{c}^m$ are used as initial values for the coefficients $\acc{\ve{c}_{\ve{\alpha}}: \ve{\alpha}\in \caa_m}$ of the stochastic surrogate in the optimization procedure, i.e., $\ve{c}_{\ve{\alpha}}$ for $\ve{\alpha}\in \caa_m$.

For the other coefficients $\acc{c_{\ve{\alpha}}:\ve{\alpha} \in \caa\setminus\caa_m}$, we randomly initialize their value. 

\subsection{Warm-start strategy}
Because of the form of the likelihood \cref{eq:likelihood}, the gradient at the starting point can take extremely large values when $\sigma$ is small. In this case, the optimization algorithm may become unstable and converge to an undesired local optimum. To guide the optimization, we propose a warm-start strategy summarized in \Cref{alg:WS}. We generate a decreasing sequence $\ve{\sigma} = \acc{\sigma_{1}, \ldots, \sigma_{N_s}}$ with $\sigma_{N_s} = \sigma$ (the target value). In this paper, we choose the maximum value $\sigma_{1}$ of the sequence as the square root of the \emph{leave-one-out error} $\varepsilon_{\rm LOO}$ in the mean fitting procedure (see \Cref{sec:upper} for the explanation of this choice). Then, $\ve{\sigma}$ is generated equally-spaced in the log-space between $\sqrt{\varepsilon_{\rm LOO}}$ and $\sigma$. Starting with $\sigma_{1}$ which is the largest element of $\ve{\sigma}$, we build a stochastic PCE based on \cref{eq:MLE} with the initial values defined above (the mean function estimation and random initialization). Then, the results are used as a starting point for the construction of the surrogate for $\sigma_2$. We repeat this procedure sequentially for each element in $\ve{\sigma}$ with each new starting point being the results of the previous optimization. Because the standard deviation decreases progressively to the target value and the starting point is updated accordingly, the associated gradient for each optimization prevents extremely big values.

\begin{algorithm}[h]
	\caption{Warm-start approach for estimating $\ve{c}$ with known $\sigma$}
	\label{alg:WS}
	\begin{algorithmic}[1]
		\REQUIRE $\left(\cx,\ve{y}\right)$, $\sigma$, $\caa$
		\ENSURE Coefficients $\hat{\ve{c}}$
		\STATE $\ve{c}^m, \varepsilon_{\rm LOO} \gets \OLS(\cx,\ve{y},\caa_m)$ \hfill {\% Estimation of the coefficients of the mean function}
		\STATE $c^0_{\ve{\alpha}} \gets c^m_{\ve{\alpha}}$ for $\ve{\alpha}\in \caa_m$ and randomly initialize $\acc{{c^0_{\ve{\alpha}}: \ve{\alpha}\in \caa \setminus \caa_m}}$
		\STATE $\ve{\sigma}_{\log} \gets {\rm linspace}\left(\log\left(\sqrt{\varepsilon_{\rm LOO}}\right),\log(\sigma),N_s\right)$
		\STATE $\ve{\sigma} \gets \exp\left(\ve{\sigma}_{\log}\right)$
		\FOR{$i \gets 1,\ldots,N_s$}
		\STATE Solve \cref{eq:MLE} to compute $\ve{c}^i$ using $\ve{c}^{i-1}$ as initial values
		\ENDFOR
		\STATE $\hat{\ve{c}} \gets \ve{c}^{N_s}$
	\end{algorithmic}
\end{algorithm}

\subsection{Cross-validation}
\label{sec:cv}
As explained in \Cref{sec:discuss}, the hyperparameter $\sigma$ cannot be jointly estimated together with the PCE coefficients $\ve{c}$ because the likelihood function can reach $+\infty$ for certain choices of $\ve{c}$ and $\sigma = 0$. As a result, $\sigma$ should be tuned separately from the estimation of $\ve{c}$. 

In this paper, we propose applying cross-validation (CV) \cite{Hastie2001} to selecting the optimal value of $\sigma$. More precisely, the data $(\cx,\ve{y})$ are randomly partitioned into $N_{\rm cv}$ equal-sized groups $\acc{V_k: k=1,\ldots,N_{\rm cv}}$ (so-called $N_{\rm cv}$-fold CV). For $k \in \acc{1, \ldots,N_{\rm cv}}$, we pick the $k$-th group $V_k$ as the validation set and the other $N_{\rm cv}-1$ folds denoted by $V_{{\sim} k}$ as the training set. The latter is used to build a stochastic PCE following \cref{eq:MLE} and \cref{alg:WS}, which yields
\begin{equation}\label{eq:MLEcv}
	\hat{\ve{c}}_k(\sigma) = \arg\max_{\ve{c}} \; \sum_{i \in V_{{\sim} k}}
	\log\left(\tilde{l}\left(\ve{c};\ve{x}^{(i)},y^{(i)},\sigma\right)\right).
\end{equation}
Note that the coefficients depend on the value of $\sigma$, and thus we explicitly write them as functions of $\sigma$. The validation set $V_k$ is then used to evaluate the \emph{out-of-sample} performance:
\begin{equation}
	\mathrm{l}_{k}(\sigma) = \sum_{i \in V_k} \log\left(\tilde{l}\left(\hat{\ve{c}}_k(\sigma);\ve{x}^{(i)},y^{(i)},\sigma\right)\right).
\end{equation}
We repeat this procedure for each group of the partition $\acc{V_k: k=1,\ldots,N_{\rm cv}}$ and sum up the respective score to estimate the generalized performance, referred to as \emph{CV score} in the sequel. Then, the optimal value of $\sigma$ is selected as the one that maximizes this CV score:
\begin{equation}\label{eq:cv}
	\hat{\sigma} = \arg\max_{\sigma} \sum_{k=1}^{N_{\rm cv}} \mathrm{l}_{k}(\sigma).
\end{equation}
Because of the nested optimization in \cref{eq:MLEcv}, the gradient of \cref{eq:cv} is difficult to derive. In this paper, we apply the derivative-free Bayesian optimizer \cite{Snoek2012} to solving \cref{eq:cv} and search for $\sigma$ within the range $[0.1,1]\times\sqrt{\varepsilon_{\rm LOO}}$. The upper bound of the interval is explained in \Cref{sec:upper}. The lower bound is introduced to prevent numerical instabilities near $\sigma = 0$. According to our investigations, the optimal value $\hat{\sigma}$ is always within the proposed interval. 

After solving \cref{eq:cv}, the selected $\hat{\sigma}$ is used in \cref{eq:MLE} with all the available data to build the final surrogate. 

Large value of $N_{\rm cv}$ can lead to high computational cost, especially when $N$ is big. In this paper, we choose $N_{\rm cv} = 10$ for $N<200$ (small data set), $N_{\rm cv}=5$ for $200\leq N<1{,}000$ (moderate data set) and $N_{\rm cv} = 3$ for $N\geq 1{,}000$ (big data set).

\subsection{Adaptivity}
\label{sec:adapt}
The method developed in \Cref{sec:MLE,sec:cv} allows us to build a stochastic PCE for a given distribution of the latent variable $Z$ and truncated set $\caa$ of polynomial chaos basis. In principle, one can choose any continuous probability distribution for the latent variable and a large truncated set. However, in practice, certain types of latent variables may require a lot of basis functions to  approximate well the shape of the response distribution. This leads to many model parameters to estimate, which would cause overfitting when only a few data are available. In this section, we propose a procedure to iteratively find a suitable distribution for the latent variable $Z$ and truncation scheme $\caa$.

We consider $N_z$ candidate distributions $\ve{D}=\acc{D_1,\ldots,D_{N_z}}$ for the latent variable, $N_p$ degrees $\ve{p} = \acc{p_1,\ldots,p_{N_p}}$ and $N_q$ $q$-norms $\ve{q} = \acc{p_1,\ldots,p_{N_p}}$ that are used to define the hyperbolic truncation scheme in \cref{eq:qnorm}. Both $\ve{p}$ and $\ve{q}$ are sorted in increasing order. 

The adaptive procedure is shown in \Cref{alg:adapt} and described here. For each type of latent variable and truncation set $\caa = \caa^{p,q,M}$, we first apply the hybrid LAR algorithm developed by Blatman and Sudret \cite{SudretJCP2011} to fitting the mean function $\hat{m}(\ve{x})$ as shown in \cref{eq:meanfit}. This algorithm only selects the most important basis among the candidate set $\caa_m$ defined in \cref{eq:cam}. To reduce the total number of unknowns in the optimization \cref{eq:MLE}, we exclude from $\caa$ the basis functions in $\caa_m$ that are not selected by hybrid LAR. In other words, we only estimate the coefficients associated with the basis functions that either have $\alpha_{z} \neq 0$ or are selected by the hybrid LAR when fitting the mean function $m(\ve{x})$. Then, we use the methods presented in \Cref{sec:MLE,sec:cv} to build a stochastic PCE for $\caa$ and record the CV score. The latter is used for model comparisons, and the one with the best CV score is selected as the final surrogate.

\begin{algorithm}[h]
	\caption{Adaptive algorithm for building a stochastic PCE}
	\label{alg:adapt}
	\begin{algorithmic}[1]
		\REQUIRE $\left(\cx,\ve{y}\right)$, $\ve{D}$, $\ve{p}$, $\ve{q}$
		\ENSURE $D_{opt}$, $\caa_{opt}$, $\hat{\ve{c}}$, $\hat{\sigma}$
		\STATE $\mathrm{l}_{opt} \gets -\infty$
		\FOR{$i_{z} \gets 1,\ldots,N_z$}
		\STATE Set $Z\sim D_{i_z}$
		\FOR{$i_{p} \gets 1,\ldots,N_p$}
		\FOR{$i_{q} \gets 1,\ldots,N_q$}
		\STATE $\caa \gets \caa^{p_{i_p},q_{i_q},M+1}$
		\STATE $\caa_m \gets \acc{\ve{\alpha}: \ve{\alpha}\in \caa, \alpha_{z} = 0}, \; \caa_c \gets \caa\setminus \caa_m$
		\STATE $\caa_n \gets \LAR\left(\cx,\ve{y},\caa_m\right)$ \hfill {\% Selection of the basis for $\hat{m}(\ve{x})$}
		\STATE $\caa \gets \caa_n \cup \caa_c$
		\STATE Apply the algorithm presented in \Cref{sec:MLE,sec:cv} to build a stochastic PCE with $\caa$, which gives $\ve{c}$, $\sigma$, and the CV score $\mathrm{l}_{i_p,i_q}$ associated with $\sigma$.
		\ENDFOR
		\ENDFOR
		\ENDFOR
		\STATE Return the model with the maximum CV score
	\end{algorithmic}
\end{algorithm}

In order to avoid going through all the possible combinations, we propose a heuristic \emph{early stopping criterion} for both degree and $q$-norm adaptivity. If two consecutive increases of $q$-norm cannot improve the CV score, the inner loop for $q$-norm adaptivity stops. Besides, if the best model (among all the $q$-norms) of a larger degree decreases the CV score, the algorithm stops exploring higher degrees. Note that the early stopping is only applied to $p$- and $q$-adaptivity, but all the candidate distributions are investigated.
%

\begin{figure}[!htbp]
	\centering
	\includegraphics[width=0.98\linewidth, keepaspectratio]{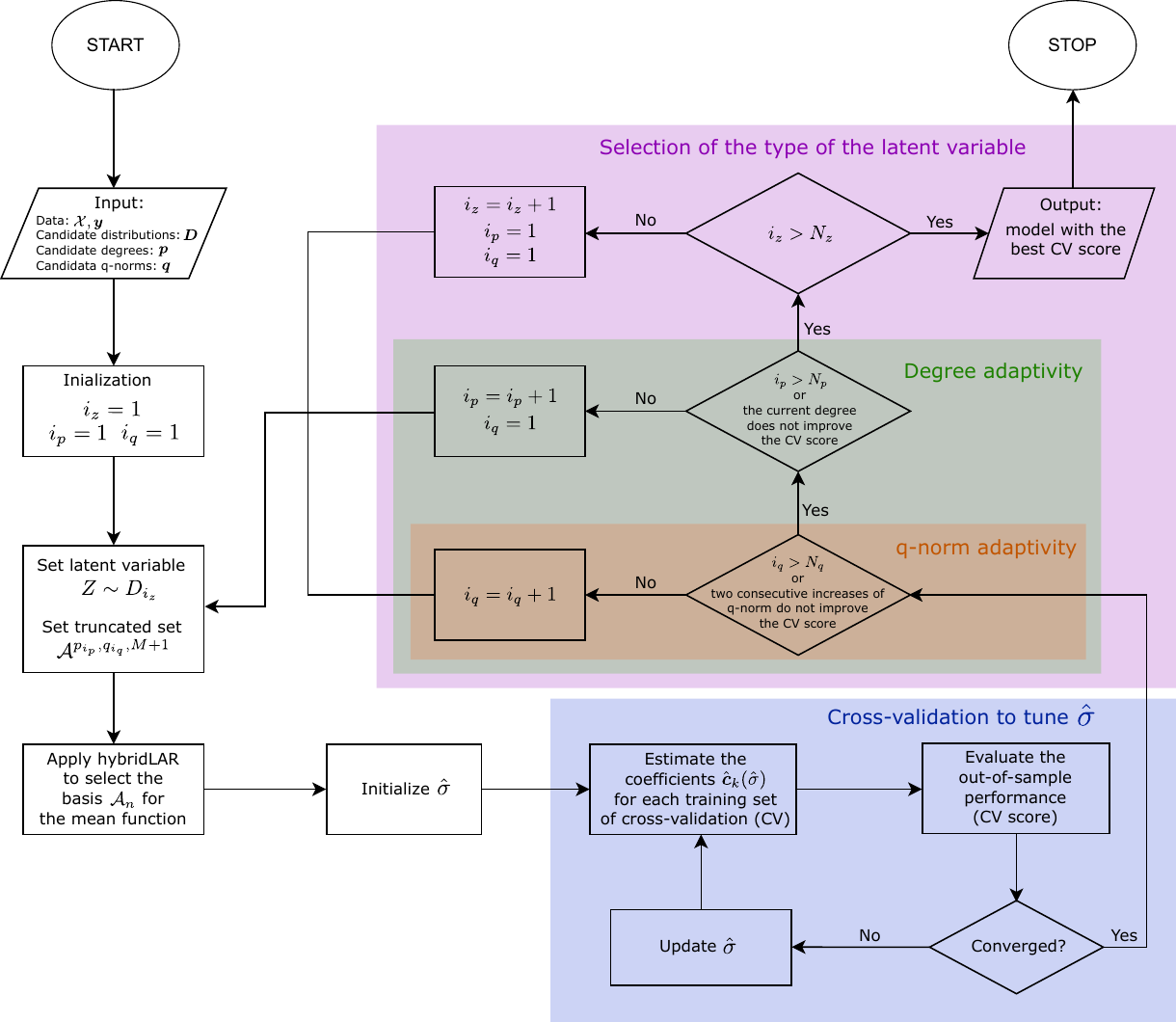}
	\caption{Flow chart of the procedure to adaptively build a stochastic PCE}
	\label{fig:flowChart}
\end{figure}

In summary, we sketch the overall procedure (presented in \Crefrange{sec:MLE}{sec:adapt}) to adaptively build a stochastic PCE from data in \Cref{fig:flowChart}.

In the application examples, we choose $N_Z=2$ possible distributions for the latent variable $Z$, namely a standard normal distribution $\cn(0,1)$ and a uniform distribution $\cu(-1,1)$. The truncation parameters $\ve{p}$ and $\ve{q}$ may be selected according to the dimensionality $M$ of the problem and the prior knowledge on the level of non-linearity. We typically use $\ve{p}=\acc{1,2,3,4,5}$ and $\ve{q}=\acc{0.5,0.75,1}$. 

\subsection{Post-processing of stochastic polynomial chaos expansions}
\label{sec:post}
In this section, we show how to post-process a stochastic PCE for various analyses. The very feature of this surrogate is that it provides a functional mapping between the input parameters $\ve{X}$, the latent variable $Z$, and the noise term $\epsilon$:
\begin{equation}\label{eq:sample}
	\tilde{Y} \eqdef \sum_{\ve{\alpha} \in \caa} c_{\ve{\alpha}} \psi_{\ve{\alpha}}\left(\ve{X},Z\right) + \epsilon,
\end{equation}
To generate realizations of $\tilde{Y}$, we simply sample $\ve{X}$, $Z$ and $\epsilon$ following their distributions and then evaluate \cref{eq:sample}. To obtain samples of $\tilde{Y}_{\ve{x}}$ for a fixed $\ve{x}$ (e.g., to plot the conditional distribution), we follow the same procedure with fixed $\ve{X} = \ve{x}$. Moreover, \cref{eq:sample} can be easily vectorized for efficient sampling.

By generating a large number of samples, one can display the distribution of $\tilde{Y}$ and $\tilde{Y}_{\ve{x}}$ using histograms or kernel density estimation. We can also use the quadrature version in \cref{eq:quadrature} to get an explicit form of the conditional response distribution of $\tilde{Y}_{\ve{x}}$. 

In addition, because the proposed surrogate model is derived based on PCE, it inherits all the good properties of PCE. In particular, some important quantities can be directly computed by post-processing the PCE coefficients $\ve{c}$ and the parameter $\sigma$, without the need for sampling. Indeed, the mean and variance of $\tilde{Y}$ are given by
\begin{equation}
	\Esp{\tilde{Y}} = c_{\ve{0}}, \quad \Var{\tilde{Y}} = \sum_{\ve{\alpha}\in \caa\setminus{\ve{0}}}c_{\ve{\alpha}}^2 + \sigma^2.
\end{equation}
where $c_{\ve{0}}$ is the coefficient of the constant function. 

As already shown in \cref{eq:mean}, for a given value of $\ve{x}$, the mean of the model response $\tilde{Y}_{\ve{x}}$ can be computed as
\begin{equation}
	\Esp{\tilde{Y}_{\ve{x}}} = \sum_{\ve{\alpha} \in \caa, \alpha_{z} = 0 }c_{\ve{\alpha}} \psi_{\ve{\alpha}}(\ve{x}),
\end{equation}
Similarly, we can compute the variance as follows:
\begin{equation}\label{eq:varYx}
	\Var{\tilde{Y}_{\ve{x}}} = \Vare{Z,\epsilon}{\sum_{\ve{\alpha} \in \caa} c_{\ve{\alpha}} \psi_{\ve{\alpha}}\left(\ve{x},Z\right) + \epsilon} = \sum_{\ve{\alpha} \in \caa\setminus\caa_m} c^2_{\ve{\alpha}} \psi^2_{\ve{\alpha}}(\ve{x}) + \sigma^2.
\end{equation}

\subsection{Global sensitivity analysis}
In the context of global sensitivity analysis of stochastic simulators \cite{ZhuRESS2021}, various types of Sobol' indices can also be computed analytically for the proposed surrogate model. The \emph{classical Sobol' indices} are defined from the Sobol'-Hoeffding decomposition of the deterministic model given by the stochastic simulator with both the well-defined input variables $\ve{X}$ and its intrinsic stochasticity as explicit inputs $\omega$, see \cref{eq:defsto}. Since the surrogate model in \cref{eq:sample} is also a deterministic function of $\ve{X}$ and the additional variables $Z$ and $\epsilon$, the Sobol' indices can be efficiently computed from the PCE coefficients, similarly to the classical PCE-based Sobol' indices \cite{SudretRESS2008b}. For example, the first-order classical Sobol' index of the $i$-th input $X_i$ is given by
\begin{equation}\label{eq:ClassSobol}
	S_{i} \eqdef \frac{\Var{\Esp{\tilde{Y} \mid X_i}}}{\Var{\tilde{Y}}} = \frac{\sum\limits_{\ve{\alpha} \in \caa_i}  c_{\ve{\alpha}}^2 }{\sum\limits_{\ve{\alpha}\in \caa\setminus{\ve{0}}}c_{\ve{\alpha}}^2 + \sigma^2},
\end{equation}
where $\caa_i \eqdef \acc{\ve{\alpha} \in \caa: \alpha_i \neq 0,\, \alpha_j = 0\,, \forall j\neq i}$. Similarly, one can also calculate higher-order and total Sobol' indices of the model \cref{eq:sample}. Let us split the input vector into two subsets $\ve{X} = (\ve{X}_{\iu},\ve{X}_{{\sim}\iu})$, where $\iu \subset \acc{1,\ldots,M}$ and ${\sim} \iu$ is the complement of $\iu$, i.e., ${\sim} \iu = \acc{1,\ldots,M} \setminus \iu$. The higher-order and total Sobol' indices, denoted by $S_{\iu}$ and $S_{T_i}$, respectively, are given by
\begin{equation}
	S_{\iu} = \frac{\sum\limits_{\ve{\alpha} \in \caa_{\iu}}  c_{\ve{\alpha}}^2 }{\sum\limits_{\ve{\alpha}\in \caa\setminus{\ve{0}}}c_{\ve{\alpha}}^2 + \sigma^2}, \quad S_{T_i} = \frac{\sum\limits_{\ve{\alpha} \in \caa, \alpha_i\neq0}  c_{\ve{\alpha}}^2 }{\sum\limits_{\ve{\alpha}\in \caa\setminus{\ve{0}}}c_{\ve{\alpha}}^2 + \sigma^2},
\end{equation}
where $\caa_{\iu} \eqdef \acc{\ve{\alpha} \in \caa: \alpha_i \neq 0,\, \alpha_j = 0\,,\alpha_z = 0\,, \forall i\in \iu, \forall j \in {\sim} \iu}$. However, as mentioned in \cref{sec:SPCE}, the surrogate model aims only at emulating the response distribution of the simulator instead of representing the detailed data generation process. Therefore, the indices involving the artificial variables introduced in the surrogate (i.e., $Z$ and $\epsilon$), e.g., the first-order Sobol' index for $Z$ and the total Sobol' index for each component of $\ve{X}$, do not reveal the nature of the original model \cite{ZhuRESS2021}. 

The QoI-based Sobol' indices quantify the influence of the input variables on some quantity of interest of the random model response, e.g., mean, variance, and quantiles \cite{ZhuRESS2021}. As the mean function in \cref{eq:mean} is a PCE, the associated Sobol' indices can be computed in a straightforward way \cite{SudretRESS2008b}. Similar to \cref{eq:ClassSobol}, the first-order index is given by
\begin{equation}\label{eq:mSobol}
	S^m_{i} \eqdef \frac{\Var{\Esp{\tilde{m}(\ve{X}) \mid X_i}}}{\Var{\tilde{m}(\ve{X})}} = \frac{\sum\limits_{\ve{\alpha} \in \caa_i}  c_{\ve{\alpha}}^2 }{\sum\limits_{\ve{\alpha}\in \caa_m\setminus{\ve{0}}}c_{\ve{\alpha}}^2 },
\end{equation}
while higher-order and total Sobol' indices of the mean function read
\begin{equation}
	S^m_{\iu} = \frac{\sum\limits_{\ve{\alpha} \in \caa_{\iu}}c_{\ve{\alpha}}^2 }{\sum\limits_{\ve{\alpha}\in \caa_m\setminus{\ve{0}}}c_{\ve{\alpha}}^2 }, \quad S^m_{T_i} = \frac{\sum\limits_{\ve{\alpha} \in \caa, \alpha_i\neq 0}c_{\ve{\alpha}}^2 }{\sum\limits_{\ve{\alpha}\in \caa_m\setminus{\ve{0}}}c_{\ve{\alpha}}^2 }.
\end{equation}
In addition, the variance function in \cref{eq:varYx} is a polynomial. The associated Sobol' indices can be computed by building another PCE to represent \cref{eq:varYx} the without error.

\section{Numerical examples}
\label{sec:examples}
In this section, we validate the proposed method on several examples, namely case studies from mathematical finance and epidemiology and a complex analytical example with bimodal response distributions. To illustrate its performance, we compare the results obtained from the stochastic polynomial chaos expansion (SPCE) with two state-of-the-art models that are developed for emulating the response distribution of stochastic simulators. The first one is the generalized lambda model (GLaM). This surrogate uses the four-parameter generalized lambda distribution to approximate the response distribution of $Y_{\ve{x}}$ for any $\ve{x} \in \cd_{\ve{X}}$. The distribution parameters, as functions of the inputs, are represented by PCEs (see details in \cite{ZhuIJUQ2020,ZhuSIAMUQ2021}). The second model is based on kernel conditional density estimation (KCDE) \cite{Hayfield2008}. This method uses kernel density estimation to fit the joint distribution $\hat{f}_{\ve{X},Y}(\ve{x},y)$ and the marginal distribution $\hat{f}_{\ve{X}} (\ve{x})$. The response distribution is then estimated by
\begin{equation}\label{eq:KCDE}
	f_{Y\mid \ve{X}}(y\mid\ve{x}) = \frac{\hat{f}_{\ve{X},Y}(\ve{x},y)}{\hat{f}_{\ve{X}}(\ve{x})} =\frac{\sum_{i=1}^{N} \frac{1}{h_y}K_Y\left(\frac{y-y^{(i)}}{h_y}\right) \prod_{j=1}^{M} \frac{1}{h_j}K_j\left(\frac{x_j-x^{(i)}_j}{h_j}\right) }{\sum_{i=1}^{N}\prod_{j=1}^{M} \frac{1}{h_j}K_{j}\left(\frac{x_j-x^{(i)}_j}{h_{j}}\right)},
\end{equation}
where $K_y$ and $K_j$'s are the kernels for $Y$ and $X_j$'s, and $h_y$ and $h_j$'s are the associated bandwidths which are hyperparameters selected by a thorough leave-one-out cross-validation \cite{Hall2004}. 

Finally, we also consider a model where we represent the response with a normal distribution. The associated mean and variance as functions of the input $\ve{x}$ are set to the \emph{true} values obtained from the simulator. Therefore, the accuracy of such an approximation measures how close the response distribution is to the normal distribution. Moreover, this model represents the ``oracle'' of Gaussian-type mean-variance models, such as the ones presented in Marrel et al. \cite{Marrel2012} and Binois et al. \cite{Binois2018}. 

To quantitatively compare the various surrogates, we define an error metric between the simulator and the emulator by 
\begin{equation}\label{eq:Rlevel1} 
	\varepsilon = \frac{\Espe{\ve{X}}{d^2_{\rm 
				WS}\left(Y_{\ve{X}},\tilde{Y}_{\ve{X}}\right)}}{\Var{Y}},
\end{equation}
where $Y_{\ve{x}}$ is the model response, $\tilde{Y}_{\ve{x}}$ denotes
that of the surrogate (with the same input parameters as  $Y_{\ve{x}}$), and $Y$ is the model output aggregating all the uncertainties from both the input and the intrinsic stochasticity. $d_{\rm WS}$ is the \emph{Wasserstein distance of order two} \cite{Villani2008} between the two probability distributions defined by
\begin{equation}\label{eq:WS}
	d^2_{\rm WS}\left(Y_1,Y_2\right) \eqdef \Norm{Q_1 - Q_2}^2_2 =  \int_{0}^{1}\left(Q_1(u) - Q_2(u)\right)^2\D u,
\end{equation}
where $Q_1$ and $Q_2$ are the quantile functions of random variables $Y_1$ and $Y_2$, respectively. The error metric $\varepsilon$ in \cref{eq:Rlevel1} is unitless and invariant to shift and scale, i.e.,
\begin{equation}\label{eq:Rlevel2} 
	\frac{\Espe{\ve{X}}{d^2_{\rm WS}\left(aY_{\ve{X}}+b,a\tilde{Y}_{\ve{X}}+b\right)}}{\Var{aY+b}} = \frac{\Espe{\ve{X}}{d^2_{\rm WS}\left(Y_{\ve{X}},\tilde{Y}_{\ve{X}}\right)}}{\Var{Y}}.
\end{equation}

To evaluate the numerator in \cref{eq:Rlevel1}, we generate a test set $\cx_{\rm test}$ of size $N_{\rm test} = 1{,}000$ from the input distribution of $\ve{X}$. The Wasserstein distance is calculated for each point $\ve{x} \in \cx_{\rm test}$ and then averaged over $N_{\rm test}$. 

We use Latin hypercube sampling (LHS) \cite{McKay1979} to generate the experimental design and the test set. The stochastic simulator is evaluated only once for each set of input parameters, i.e., we do not use replications. To study the convergence property of the surrogates, experimental designs of various sizes are investigated. Each scenario is run 20 times with independent experimental designs to account for the statistical uncertainty in the LHS design and also in the internal stochasticity of the simulator. As a result, error estimates for each size of experimental design are represented by box plots constructed from the 20 repetitions of the full analysis. 

\subsection{Geometric Brownian motion}
\label{sec:GBM}
In the first example, we consider the \emph{Black-Scholes} model that is popular in mathematical finance~\cite{McNeil2005}
\begin{equation}\label{eq:GBM} 
	\D S_t = x_1\, S_t\,\D t + x_2\, S_t \,\D W_t.  
\end{equation}
\Cref{eq:GBM} is a stochastic differential equation used to model the evolution of a stock price $S_t$. Here, $\ve{x} = (x_1,x_2)^T$ are the input variables that describe the expected return rate and the volatility of the stock, respectively. $W_t$ is a Wiener process that represents the stochastic behavior of the market. Without loss of generality, we set the initial condition to $S_0 = 1$.

The simulator is stochastic: for a given $\ve{x}$, the stock price $S_t$ is a stochastic process, where the stochasticity comes from $W_t$. In this example, we are interested in $Y_{\ve{x}} = S_1$, which corresponds to the stock value at $t=1$ year. We set $X_1 \sim \cu(0,0.1)$ and $X_2\sim \cu(0.1,0.4)$ to represent the uncertainty in the return rate and the volatility, where the ranges are selected based on real data \cite{Reddy2016}.

The solution to \cref{eq:GBM} can be derived using It\^o calculus 
\cite{Shreve2004}: $Y_{\ve{x}}$ follows a lognormal distribution defined by
\begin{equation}\label{eq:GMB_solu}
	Y_{\ve{x}} \sim \cl\cn\left(x_1-\frac{x^2_2}{2},x_2\right).
\end{equation}
As the distribution of $Y_{\ve{x}}$ is known analytically in this simple example, we can sample directly from the response distribution to get the model output instead of simulating the whole path of $S_t$.

\begin{figure}[!htbp]
	\centering
	\begin{subfigure}[b]{.495\linewidth}
		\centering
		\includegraphics[height=0.58\linewidth, keepaspectratio]{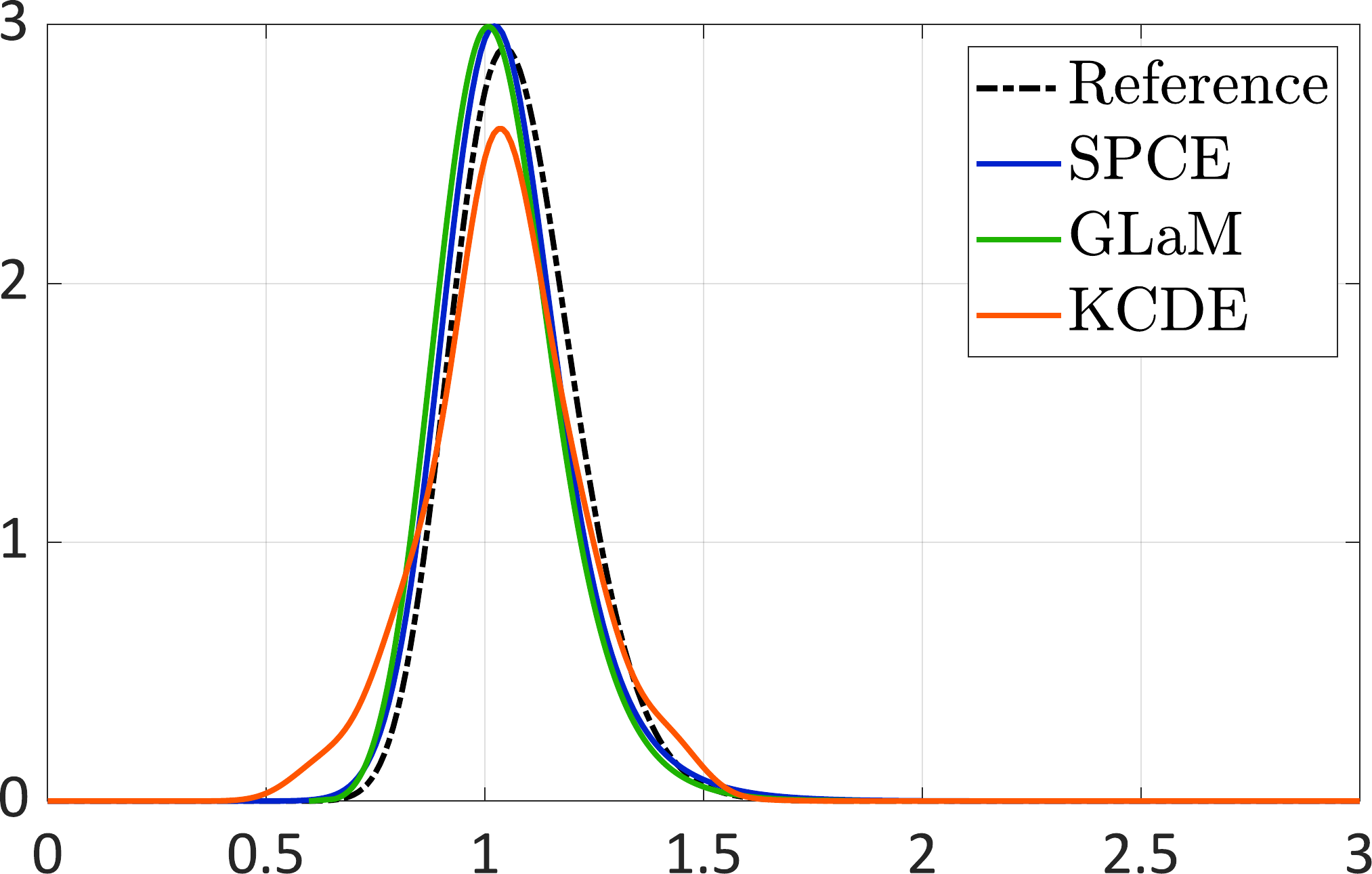}
		\caption{PDF for $\ve{x} = (0.07,0.13)^T$}
		\label{fig:GBMpdf1}
	\end{subfigure}
	\begin{subfigure}[b]{.495\linewidth}
		\centering
		\includegraphics[height=0.58\linewidth, keepaspectratio]{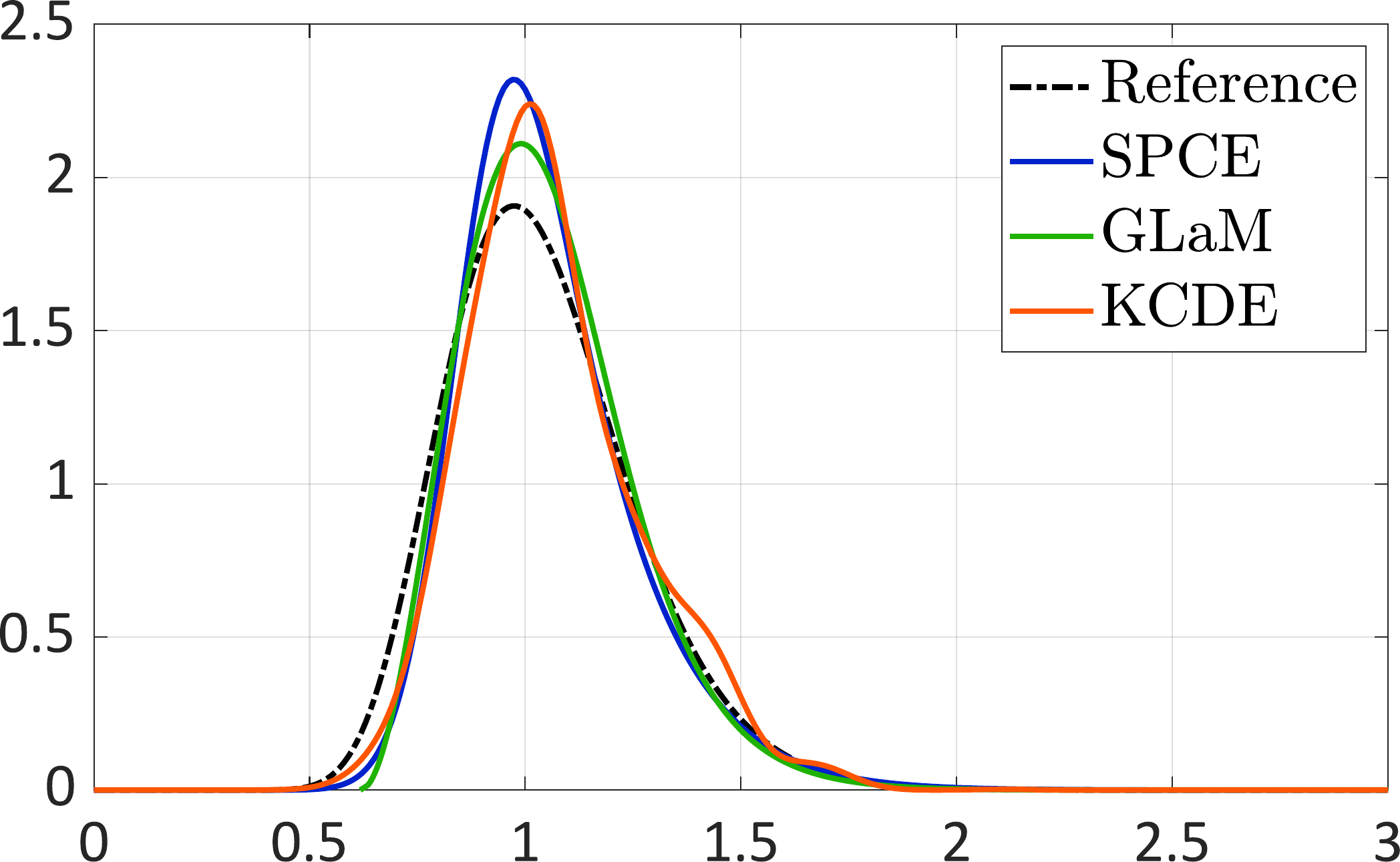}
		\caption{PDF for $\ve{x} = (0.04,0.21)^T$}
		\label{fig:GBMpdf2}
	\end{subfigure}
	\begin{subfigure}[b]{.495\linewidth}
		\centering
		\includegraphics[height=0.58\linewidth, keepaspectratio]{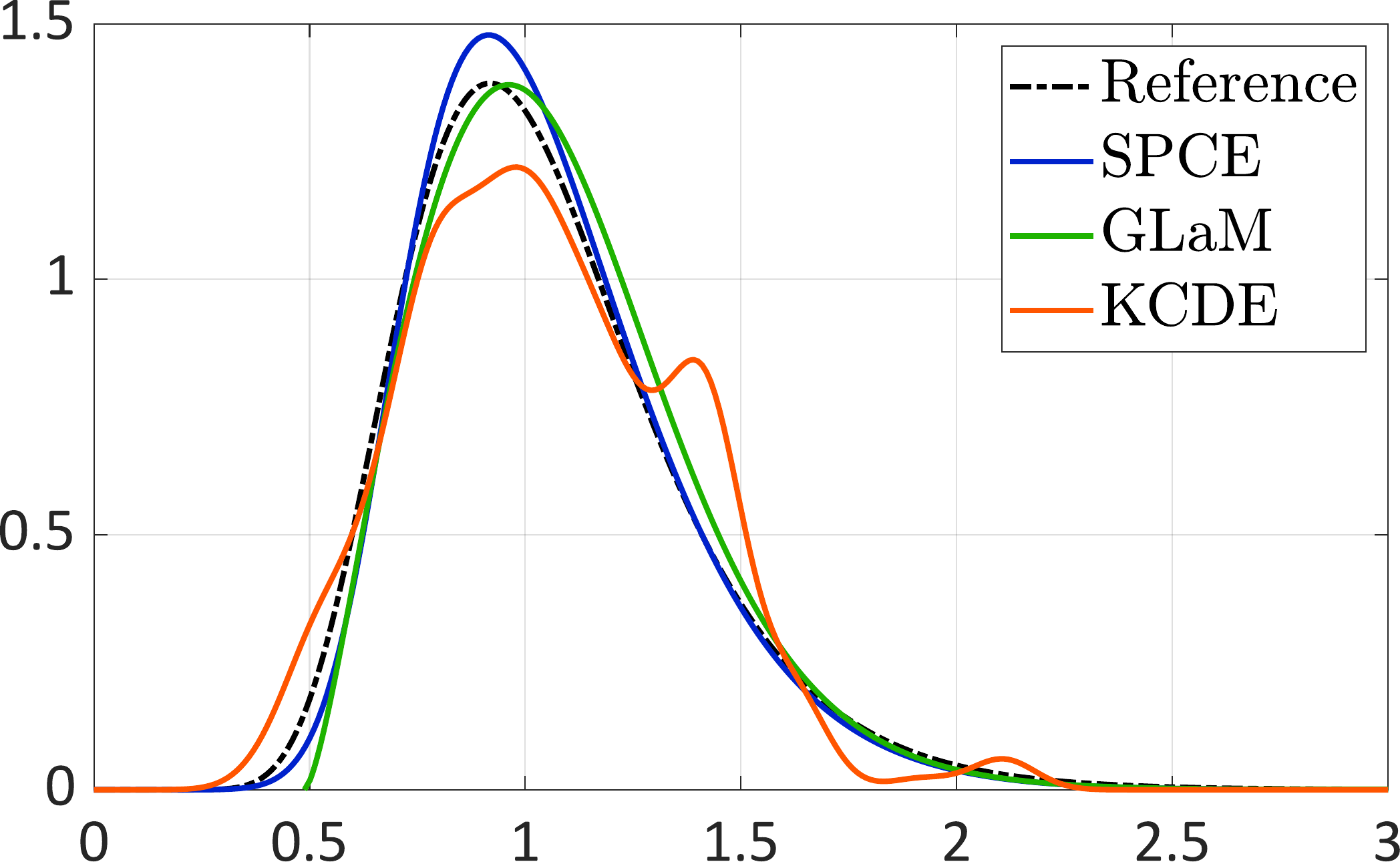}
		\caption{PDF for $\ve{x} = (0.05,0.3)^T$}
		\label{fig:GBMpdf3}
	\end{subfigure}
	\begin{subfigure}[b]{.495\linewidth}
		\centering
		\includegraphics[height=0.58\linewidth, keepaspectratio]{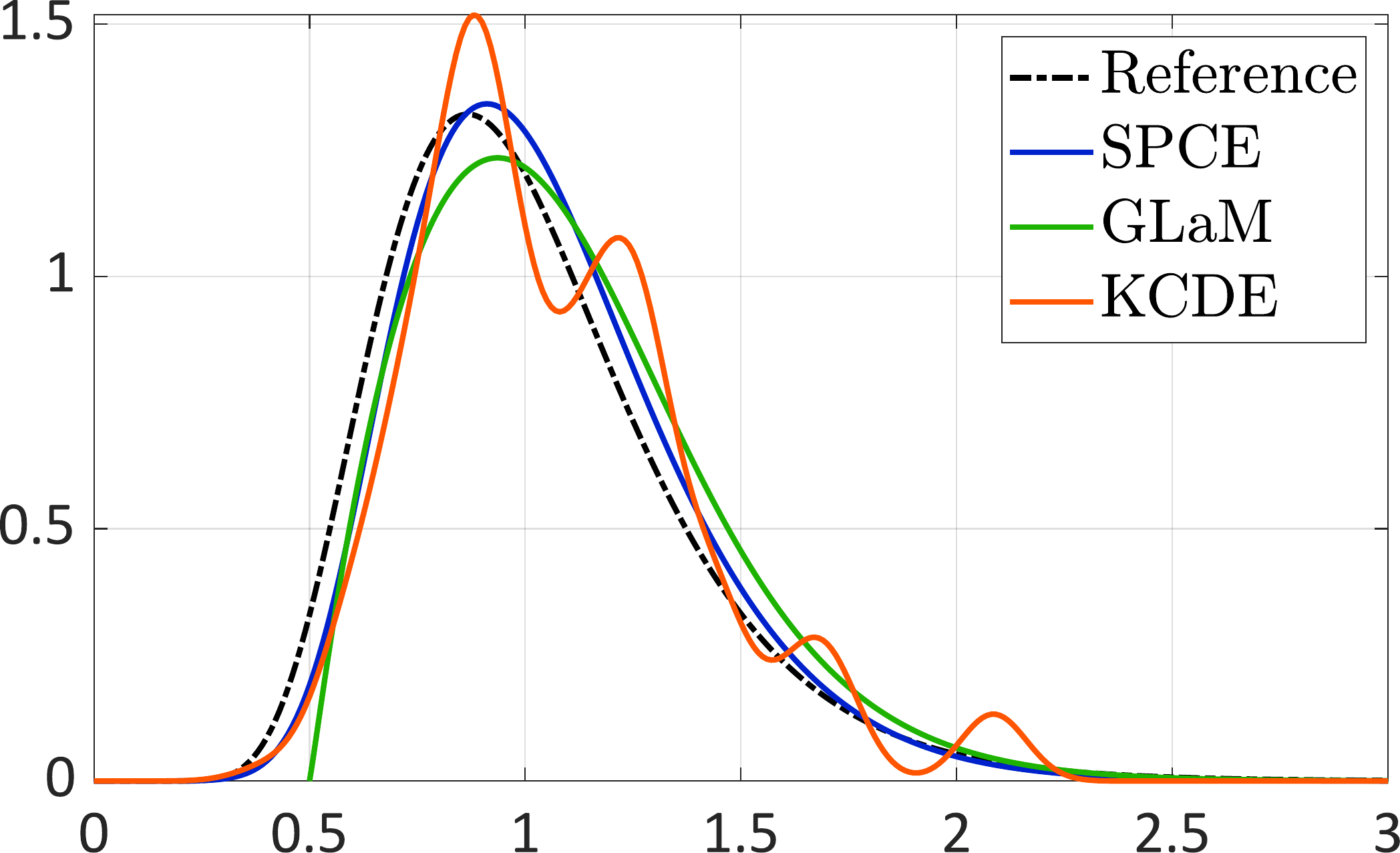}
		\caption{PDF for $\ve{x} = (0.02,0.33)^T$}
		\label{fig:GBMpdf4}
	\end{subfigure}
	\caption{Geometric Brownian motion --- Comparisons of the emulated PDFs, $N=400$.}
	\label{fig:GBMpdf}
\end{figure}

\Cref{fig:GBMpdf} illustrates four response PDFs predicted by the considered surrogates built on an experimental design of size $N=400$. We observe that with $400$ model runs, both SPCE and GLaM accurately represent the variation of the response PDF. Moreover, SPCE better represents the left tail in \cref{fig:GBMpdf4}. In contrast, KCDE can well approximate the response PDF for low volatility (in \cref{fig:GBMpdf1}) but exhibits unrealistic oscillations in the case of high volatility.

\begin{figure}[!htbp]
	\centering
	\includegraphics[width=0.45\linewidth, keepaspectratio]{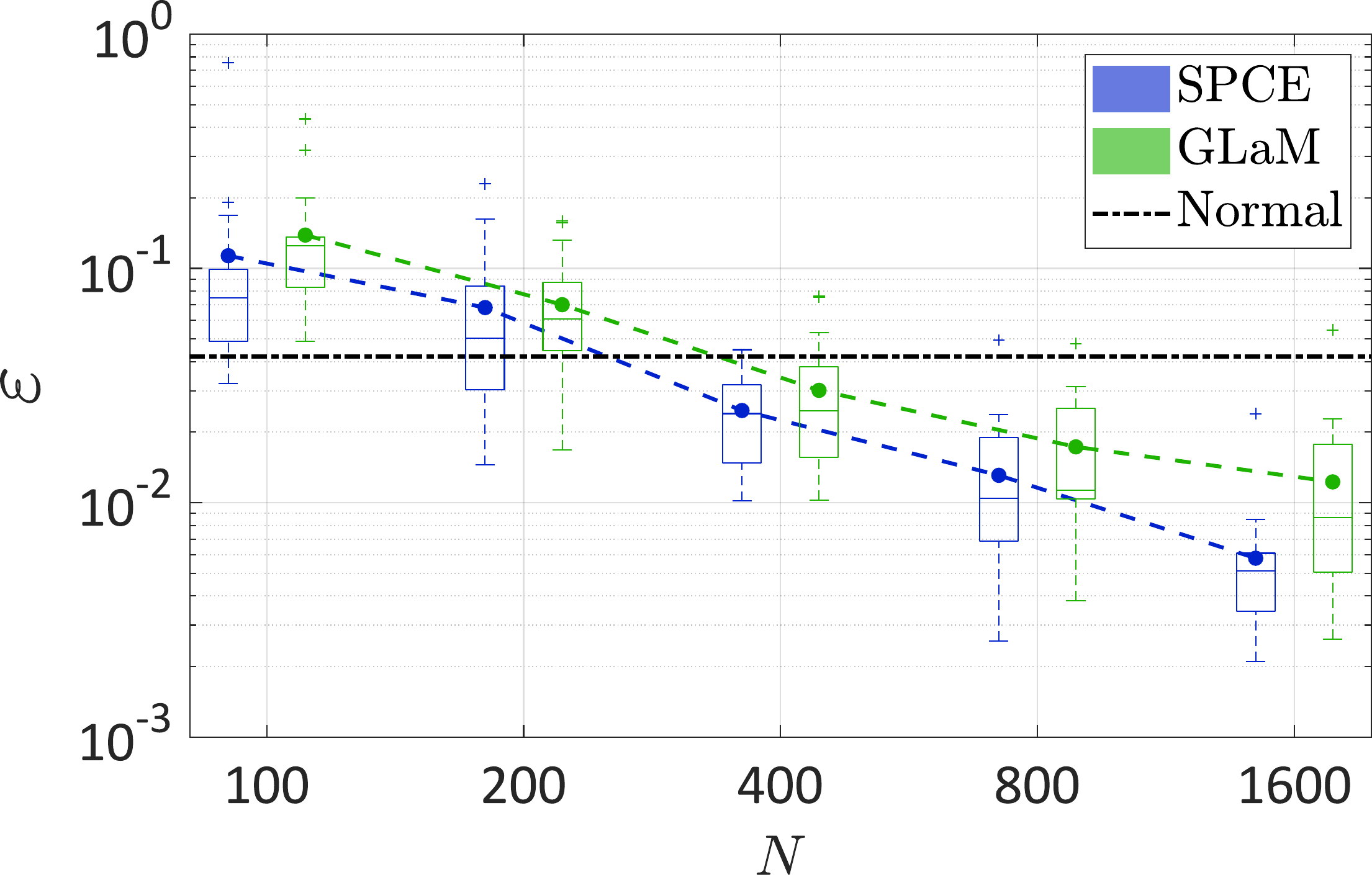}
	\hspace{5mm}
	\includegraphics[width=0.45\linewidth, keepaspectratio]{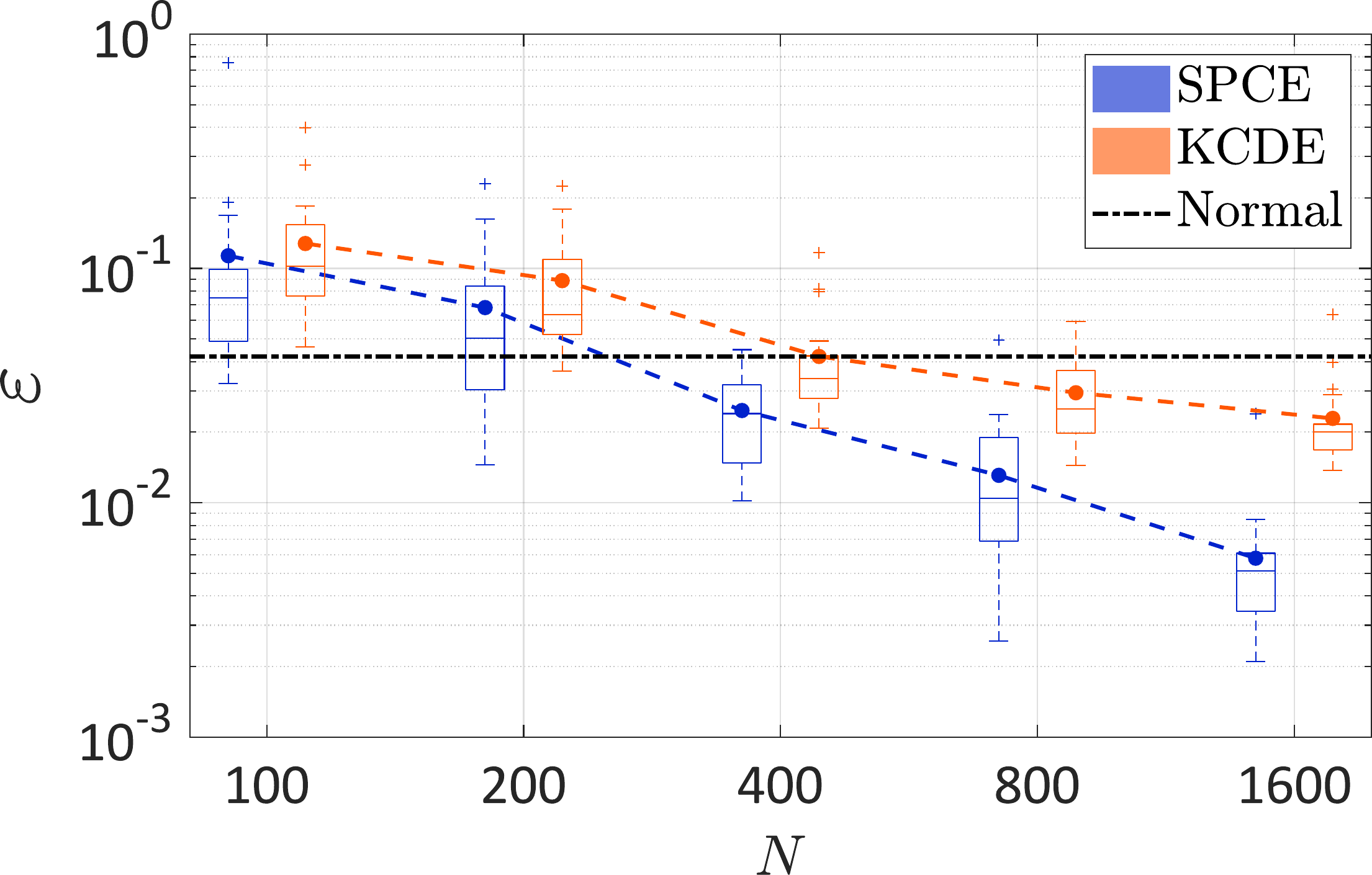}
	\caption{Geometric Brownian motion --- Comparison of the convergence of the surrogate models. The dashed lines denote the average value over 20 repetitions of the full analysis, whereas the box plot summarize the 20 results. The black dash-dotted line represents the error of the model assuming that the response distribution is normal and using the true mean and variance.}
	\label{fig:GBM_WS}
\end{figure}

For convergence studies, we vary the size of the experimental design $N \in \acc{100;200;400;800;1{,}600}$ and plot the error $\varepsilon$ defined in \cref{eq:Rlevel1} with respect to $N$ in \cref{fig:GBM_WS}. In order to show more details, each subfigure in \cref{fig:GBM_WS} compares SPCE with one competitor. We observe that the average error of KCDE built on $N=400$ model runs is similar to the best normal approximation, whereas both SPCE and GLaM provide smaller errors. Compared with KCDE and GLaM, the average performance of SPCE is always the best for all sizes of experimental design. For large $N$, namely $N=1{,}600$, the average error of SPCE is less than half of that of KCDE, and the spread of the error is narrower than that obtained by GLaM.

\subsection{Stochastic SIR model}
\label{sec:SIR}
The second example is the stochastic \emph{Susceptible-Infected-Recovered} (SIR) model frequently used in epidemiology \cite{Britton2010}. This model simulates the outbreak of an infectious disease which spreads out through stochastic contacts between infected and susceptible individuals. The simulator is a compartmental state-space model. More precisely, a population of $P$ individuals at time $t$ is partitioned into three groups: (1) \emph{susceptible individuals} who have not caught the disease and may be infected by close contact with infectious patients; (2) \emph{infected individuals} who are contaminated and infectious; (3) \emph{recovery individuals} who have recovered from the disease and are immune to future infections. The count of each group is denoted by $S_t$, $I_t$, and $R_t$, respectively. Because no newborn or death is considered, the three quantities satisfy $E_t+I_t+R_t = P$. As a result, any two out of the three counts, e.g., $E_t$ and $I_t$, can characterize the configuration of the population of size $P$ at time $t$. 

\begin{figure}[!htbp]
	\centering
	\includegraphics[width=0.65\linewidth]{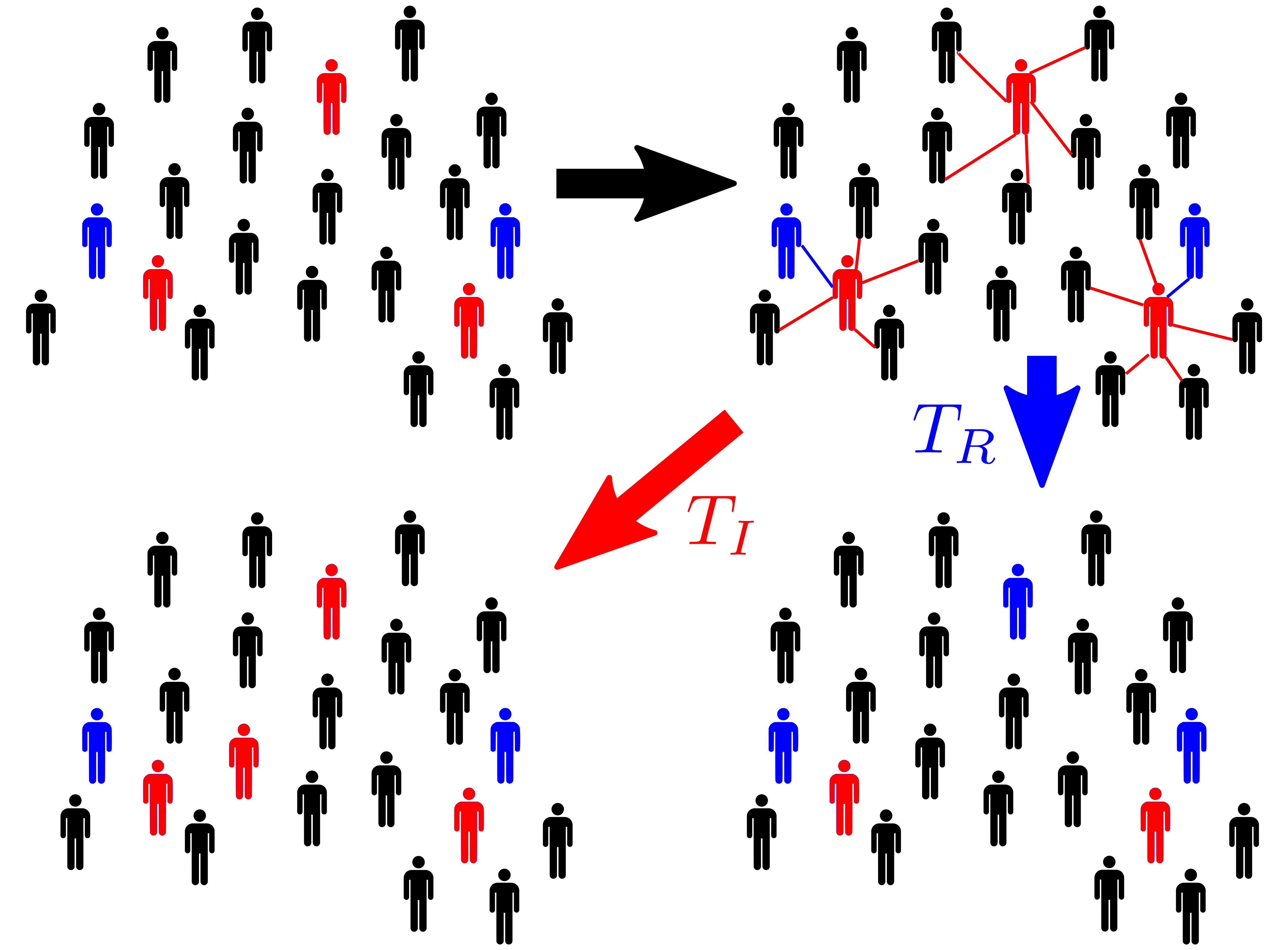}
	\caption{Dynamics of the stochastic SIR model: black icons stand for 
		susceptible individuals, red icons represent infected individuals, and blue icons are the ones that have recovered.}
	\label{fig:SIR}
\end{figure}

\Cref{fig:SIR} illustrates the dynamics of the model, where the black icons stand for susceptible individuals, the red icons correspond to infected persons, and the blue icons are the ones who have recovered. At time $t$, the state of the population is given by $(S_t,I_t)$ (the top left panel of \cref{fig:SIR}). The next configuration depends on two transition channels: infection and recovery. The first channel evolves the system to $C_I$ where one susceptible individual is infected (the bottom left panel of \cref{fig:SIR}). The recovery channel proceeds to $C_R$ where one infected person recovers (the bottom right panel of \cref{fig:SIR}). Whether the system evolves to the candidate state $C_I$ or $C_R$ depends on two random variables, $T_I$ and $T_R$ which are the respective transition time of each channel. Both $T_I$ and $T_R$ follow an exponential distribution, yet with different parameters: 
\begin{equation}\label{eq:SIRdyna}
	\begin{split}
		T_I &\sim \Exp(\lambda_I), \quad \lambda_I = \beta \frac{S_t I_t}{P}, \\
		T_R &\sim \Exp(\lambda_R), \quad \lambda_R = \gamma I_t,
	\end{split}
\end{equation}
where $\beta$ is the contact rate of an infected individual, and $\gamma$ is the recovery rate. The next configuration of the population is the one that comes first, i.e., for $T_R<T_I$, the system evolves to $C_R$ at $t+T_R$ with $S_{t+T_R}=E_t-1$ and $I_{t+T_I}=I_t+1$, and vice versa. We iterates this updating procedure until the time $T$ where $I_T = 0$ corresponding to no remaining infected individual: no infection or recovery can happen, and the outbreak stops. Since the population size is constant and recovered individuals will not be infected again, the outbreak will stop at finite time, i.e., $T< +\infty$. The simulation process described here corresponds to the \emph{Gillespie algorithm} \cite{Gillespie1977}.

The input variables of the simulator are the initial conditions $S_0$ and $I_0$ and the transitive rates $\beta$ and $\gamma$. We are interested in the total number of newly infected individuals during the outbreak without counting the initial infections, which is an important quantity in epidemics management \cite{Binois2018}. This can be calculated by the difference between the number of susceptibles at time $0$ and $T$, i.e., $Y = S_0 - S_T$. Because each updating step in \cref{eq:SIRdyna} depends on two latent variables $T_I$ and $T_R$, the simulator is stochastic. Moreover, the total number of latent variables is also random.

In this case study, we set $P=2{,}000$. To account for different scenarios, the input variables $\ve{X} = \acc{S_0,I_0,\beta,\gamma}$ are modeled as $S_0 \sim \cu(1{,}200\,,\,1{,}800)$, $I_0 \sim \cu(20,200)$, and $\beta, \gamma \sim \cu(0.5,0.75)$. The uncertainty in the first two variables is due to the lack of knowledge of the initial condition. The two transitive rates $\beta$, $\gamma$ are affected by possible interventions such as quarantine and increase of medical resources.

\begin{figure}[!htbp]
	\centering
	\begin{subfigure}[b]{.495\linewidth}
		\centering
		\includegraphics[height=0.58\linewidth, keepaspectratio]{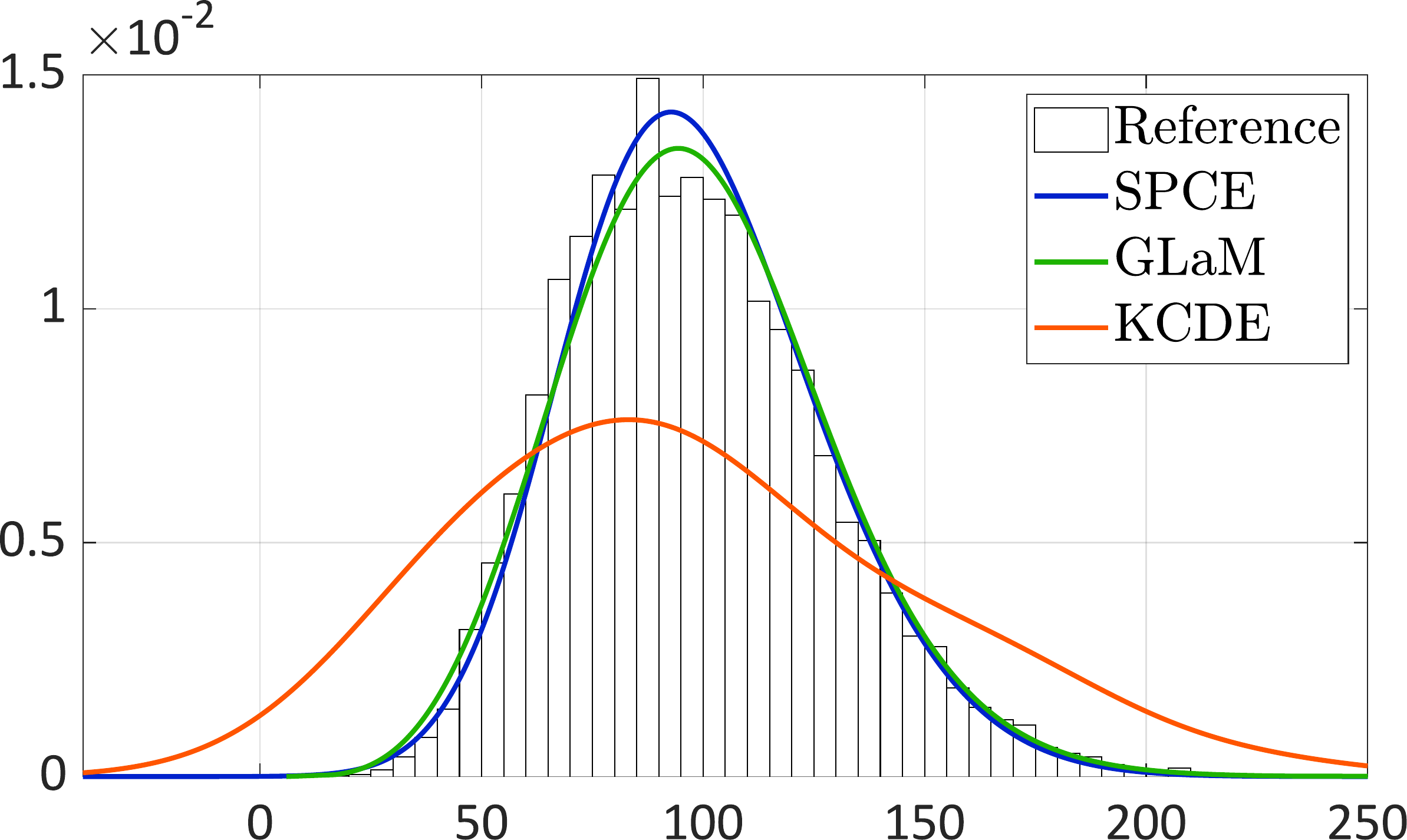}
		\caption{PDF for $\ve{x} = (1500,60,0.6,0.7)^T$}
		\label{fig:SIRpdf1}
	\end{subfigure}
	\begin{subfigure}[b]{.495\linewidth}
		\centering
		\includegraphics[height=0.58\linewidth, keepaspectratio]{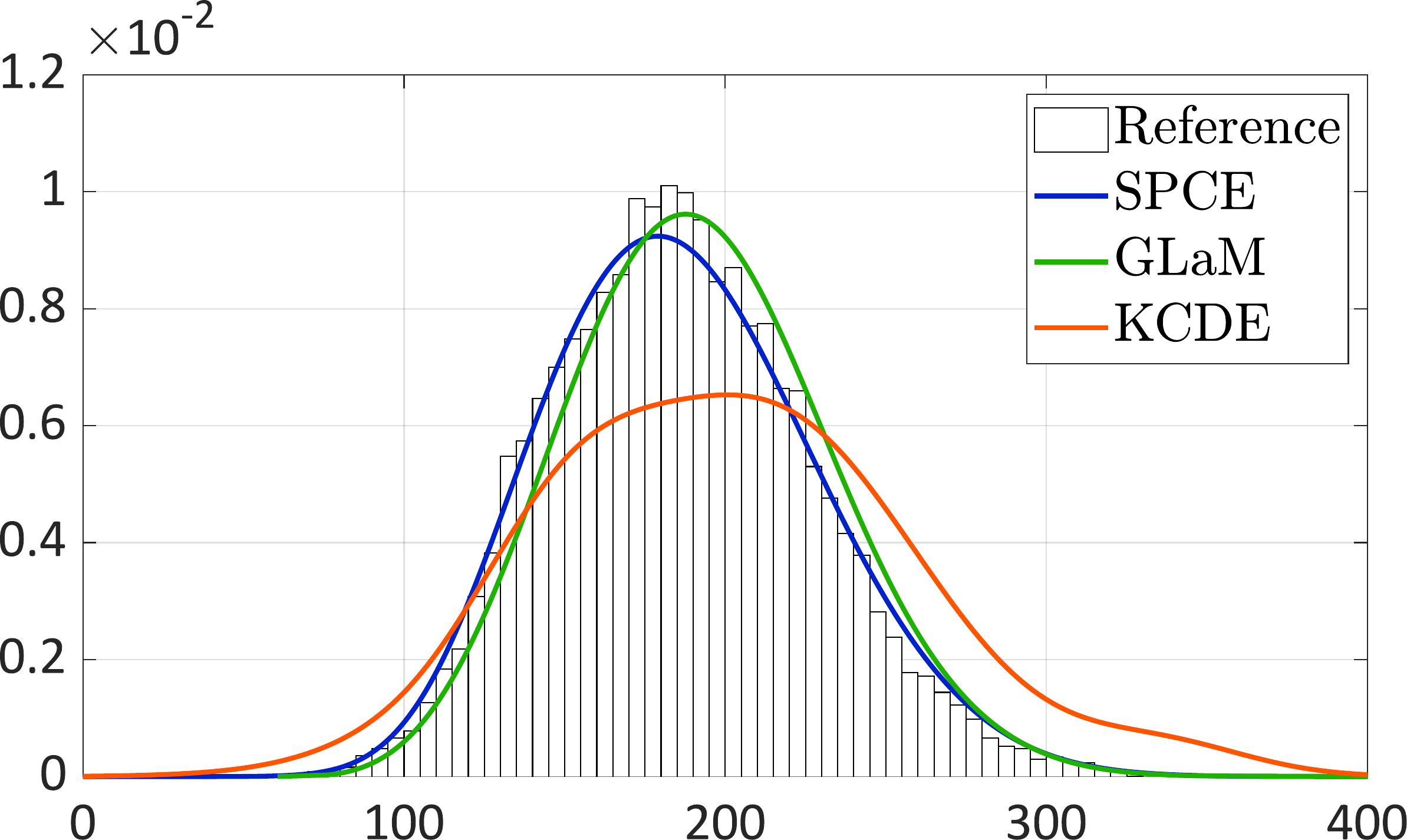}
		\caption{PDF for $\ve{x} = (1400,100,0.6,0.6)^T$}
		\label{fig:SIRpdf2}
	\end{subfigure}
	\begin{subfigure}[b]{.495\linewidth}
		\centering
		\includegraphics[height=0.58\linewidth, keepaspectratio]{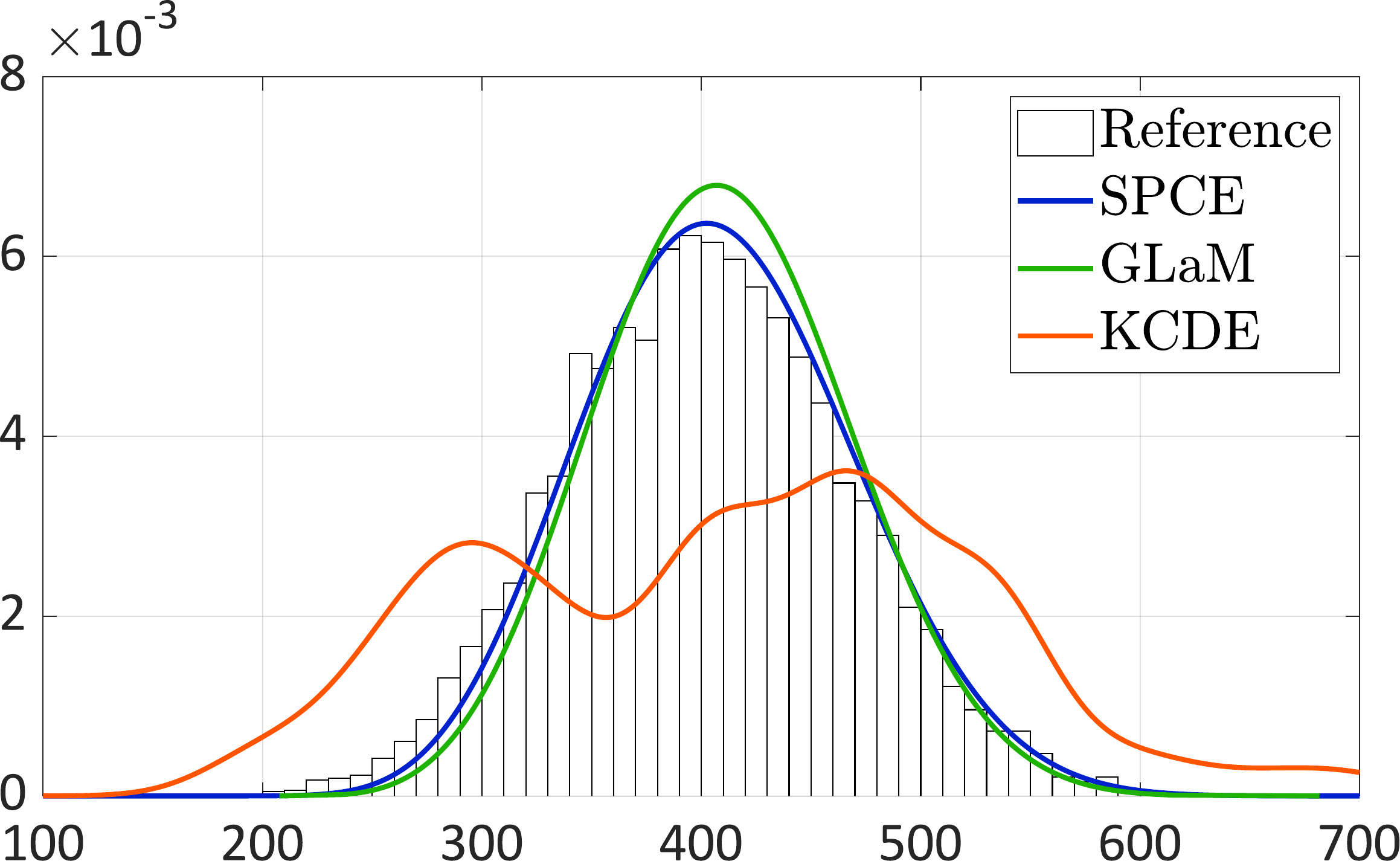}
		\caption{PDF for $\ve{x} = (1700,140,0.55,0.55)^T$}
		\label{fig:SIRpdf3}
	\end{subfigure}
	\begin{subfigure}[b]{.495\linewidth}
		\centering
		\includegraphics[height=0.58\linewidth, keepaspectratio]{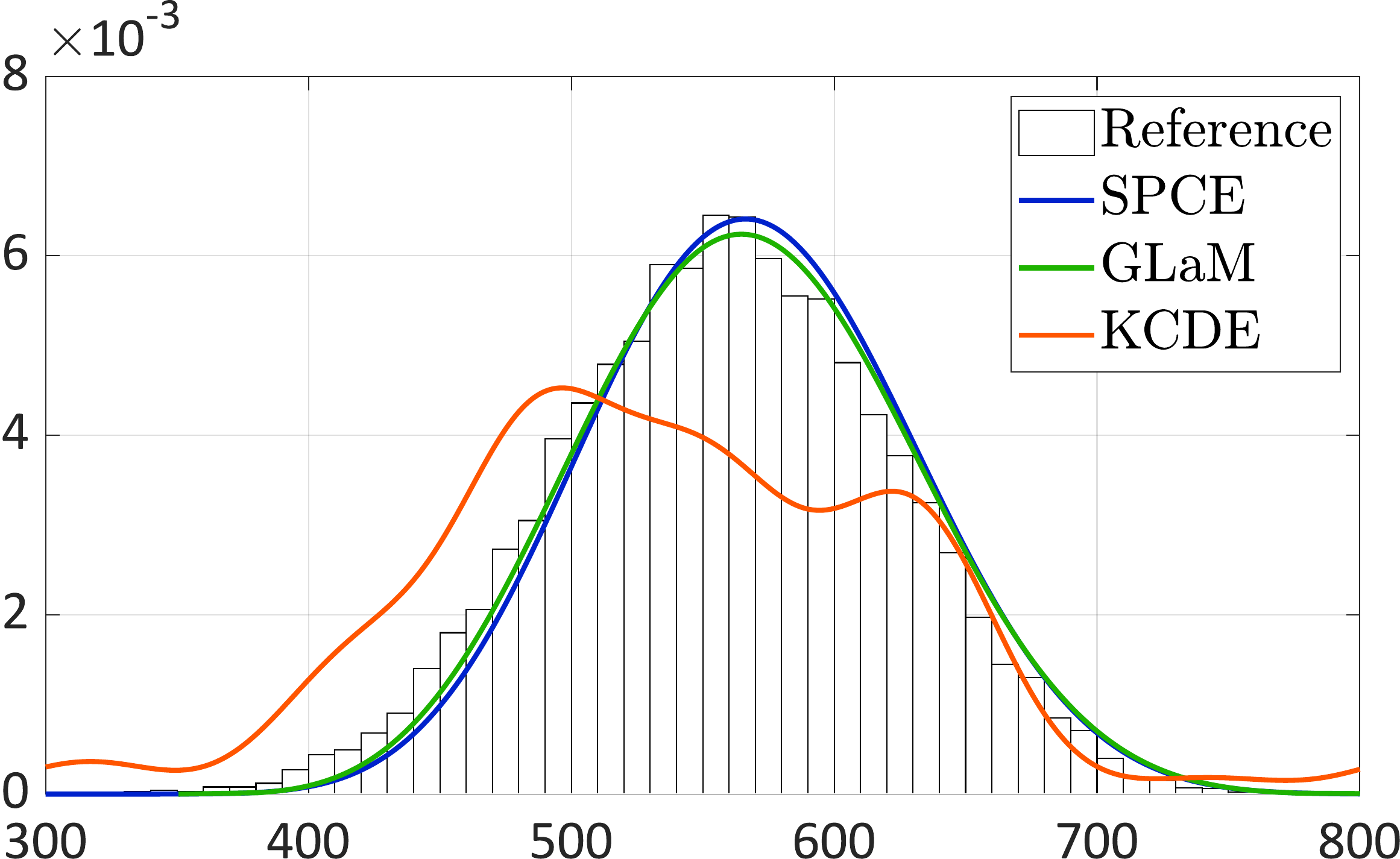}
		\caption{PDF for $\ve{x} = (1600,180,0.7,0.6)^T$}
		\label{fig:SIRpdf4}
	\end{subfigure}
	\caption{Stochastic SIR --- Comparisons of the emulated PDFs, $N=1{,}600$.}
	\label{fig:SIRpdf}
\end{figure}

\Cref{fig:SIRpdf} illustrates the response PDF for four different sets of input parameters. Because of the transition process in \cref{eq:SIRdyna}, no analytical closed-form distribution of $Y_{\ve{x}}$ can be derived. Therefore, we use $10^4$ replications for each input values to obtain the reference histograms. The surrogate models are trained on an experimental design of size $N=1{,}600$ (without any replications). We observe that the four PDFs are unimodal. The reference histogram in \cref{fig:SIRpdf1} is slightly right-skewed, while the others in \cref{fig:SIRpdf} are symmetric. SPCE and GLaM produce similar predictions of the PDF which are very close to the reference histograms. In comparison, KCDE overestimates the spread of the distributions in. Moreover, the KCDE prediction has non-negligible probability for unrealistic negative values in \cref{fig:SIRpdf1}. Besides, it exhibits relatively poor shape representations with spurious wiggles in \cref{fig:SIRpdf3} and \cref{fig:SIRpdf4}.

\begin{figure}[!htbp]
	\centering
	\includegraphics[width=0.45\linewidth, keepaspectratio]{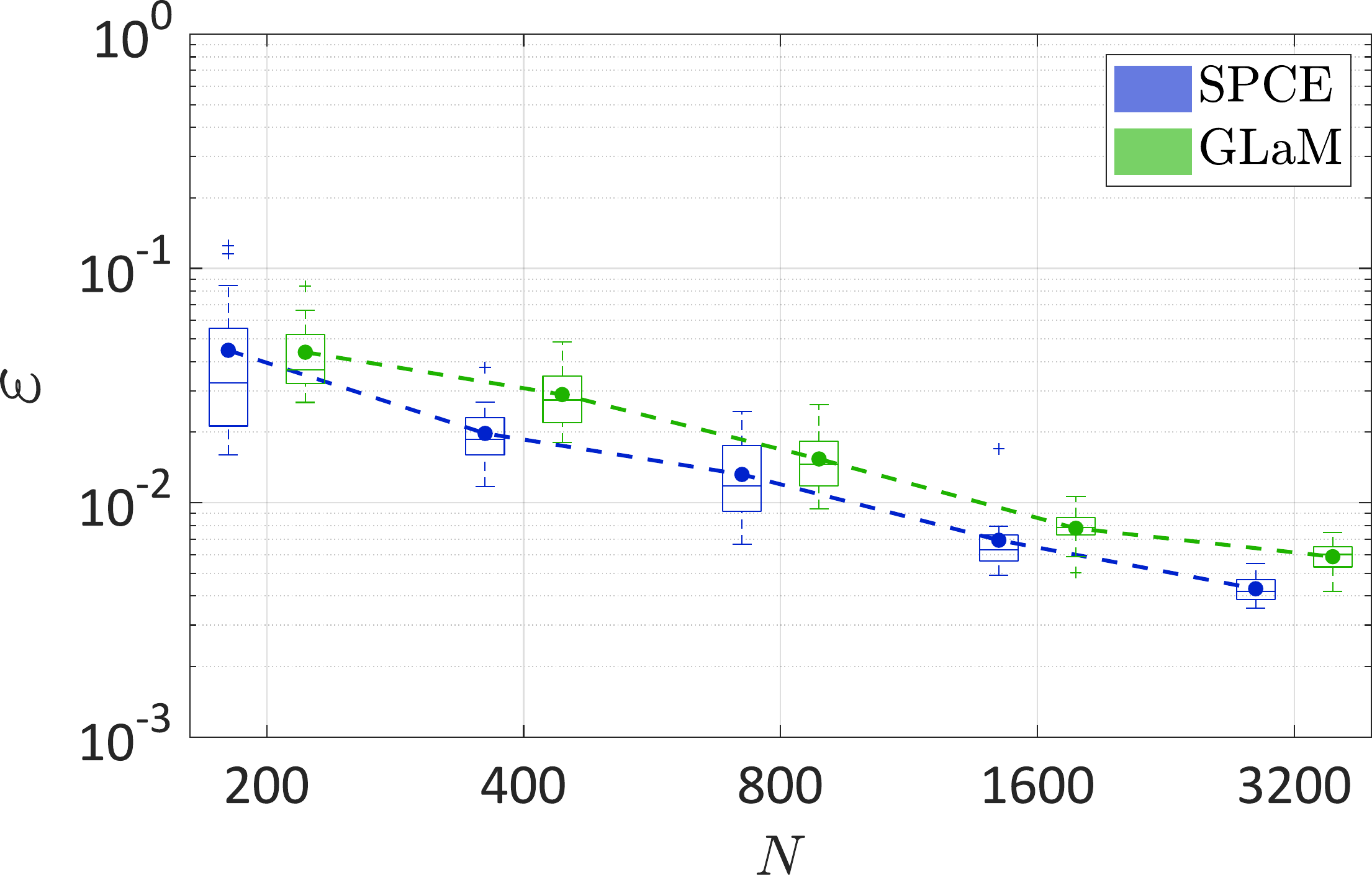}
	\hspace{5mm}
	\includegraphics[width=0.45\linewidth, keepaspectratio]{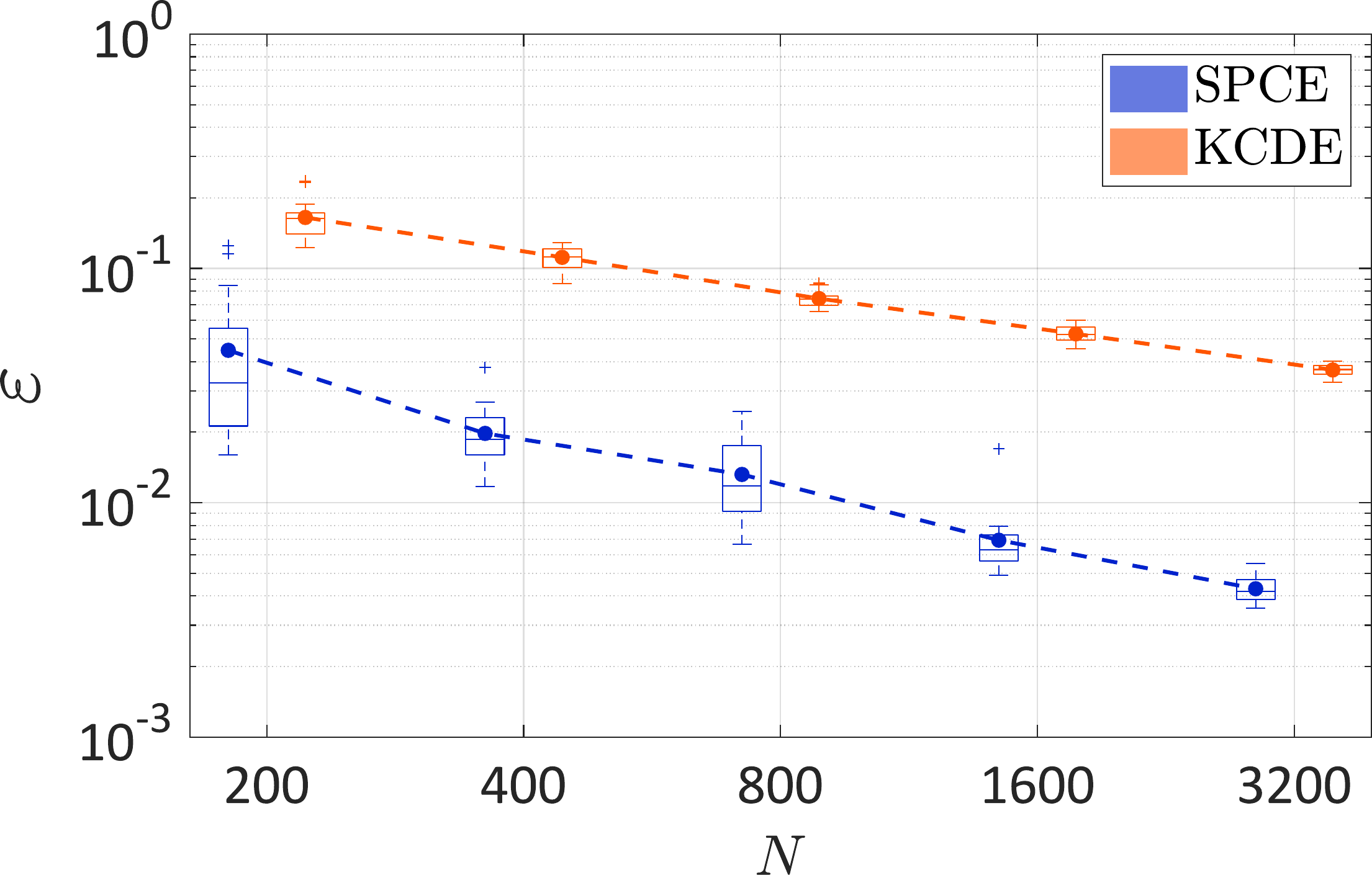}
	\caption{Stochastic SIR --- Comparison of the convergence of the surrogate models. The dashed lines denote the average value over 20 repetitions of the full analysis, whereas the box plot summarize the 20 results. The Gaussian model that assumes the response distribution being normal with the mean and variance estimated from $10^4$ replications yields an error of $6\times 10^{-4}$, which is not plotted in the figure.}
	\label{fig:SIR_WS}
\end{figure}

\Cref{fig:SIR_WS} compares the performance of the surrogates built on various sizes of experimental design $N\in \{200;\allowbreak 400;\allowbreak 800;\allowbreak 1{,}600;\allowbreak 3{,}200\}$. To evaluate the error defined in \cref{eq:Rlevel1}, the reference distribution for each $\ve{x}$ is given by the empirical distribution of $10^4$ replications. The oracle normal approximation gives an error of $6\times 10^{-4}$ which is smaller than any of the surrogates in consideration. Note that this model is not built on the training data but using the mean and variance from the $10^4$ replications for each test point. This implies that the response distribution is close to normal. We do not include this error in \cref{fig:SIR_WS} to not loose detailed comparisons of the surrogate models. \Cref{fig:SIR_WS} reveals a poor performance of KCDE in this case study. This is because the example is four-dimensional, and KCDE is a kernel-based method which is known to suffer from the \emph{curse of dimensionality}. In contrast, SPCE and GLaM are flexible parametric models, and both provide a much smaller error than KCDE for all values of $N$. Compared with GLaM, SPCE yields a similar spread of the error but demonstrates better average performance for $N \geq 400$.

\subsection{Bimodal analytical example}
\label{sec:bimodal}
\begin{figure}[!htbp]
	\centering
	\includegraphics[width=0.65\linewidth, keepaspectratio]{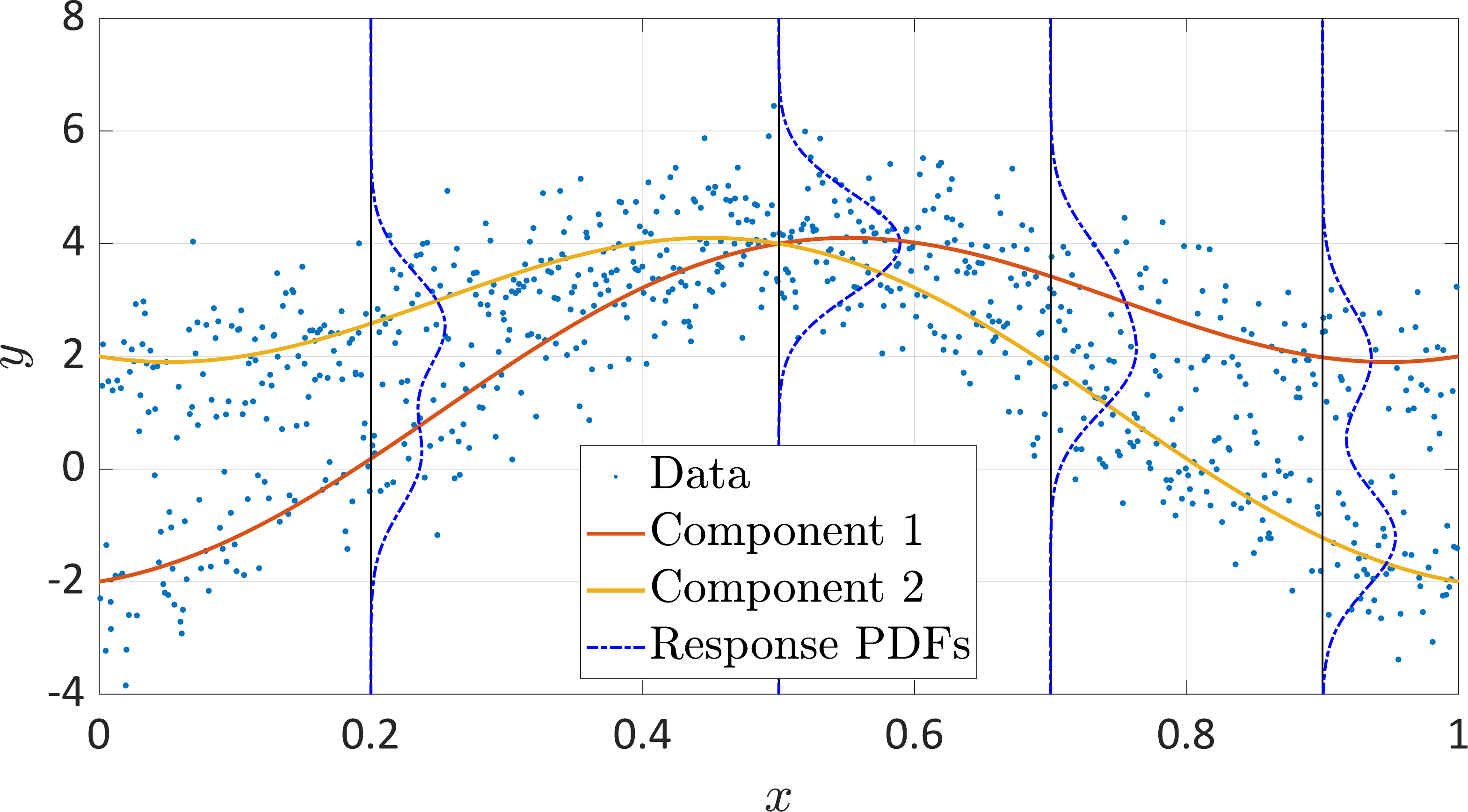}
	\caption{Bimodal analytical example --- Illustration of the model with an experimental design of $N = 800$}
	\label{fig:bimodaldata}
\end{figure}

The response distributions of the previous two examples are unimodal. In the last example, we consider a complex analytical example to test the flexibility of the stochastic polynomial chaos expansion. For this purpose, we directly define the response distribution to approximate as
\begin{equation}\label{eq:bimodal}
	f_{Y \mid X}(y \mid x) = 0.5\,\varphi\left(1.25\,y-(5\sin^2(\pi \cdot x)+5x-2.5)\right) + 0.75\,\varphi\left(1.25\,y-(5\sin^2(\pi \cdot x)-5x+2.5)\right)
\end{equation}
where $\varphi$ stands for the standard normal PDF. This response PDF is a mixture of two Gaussian PDFs with weights 0.6 and 0.8. The mean function of each component distribution depends on the input variable $x$. Let $X \sim \cu(0,1)$. With different realization of $X$, the two components change their location accordingly. \Cref{fig:bimodaldata} illustrates a data set generated by $N=800$ model runs and the mean function of each component of \cref{eq:bimodal} which varies nonlinearly with respect to the input. It is clear that the resulting conditional distribution is bimodal for small ($x \lesssim 0.2$) and large values of $x$ ($x \gtrsim 0.8$), whereas it is unimodal in between.

\begin{figure}[!htbp]
	\centering
	\begin{subfigure}[b]{.495\linewidth}
		\centering
		\includegraphics[height=0.58\linewidth, keepaspectratio]{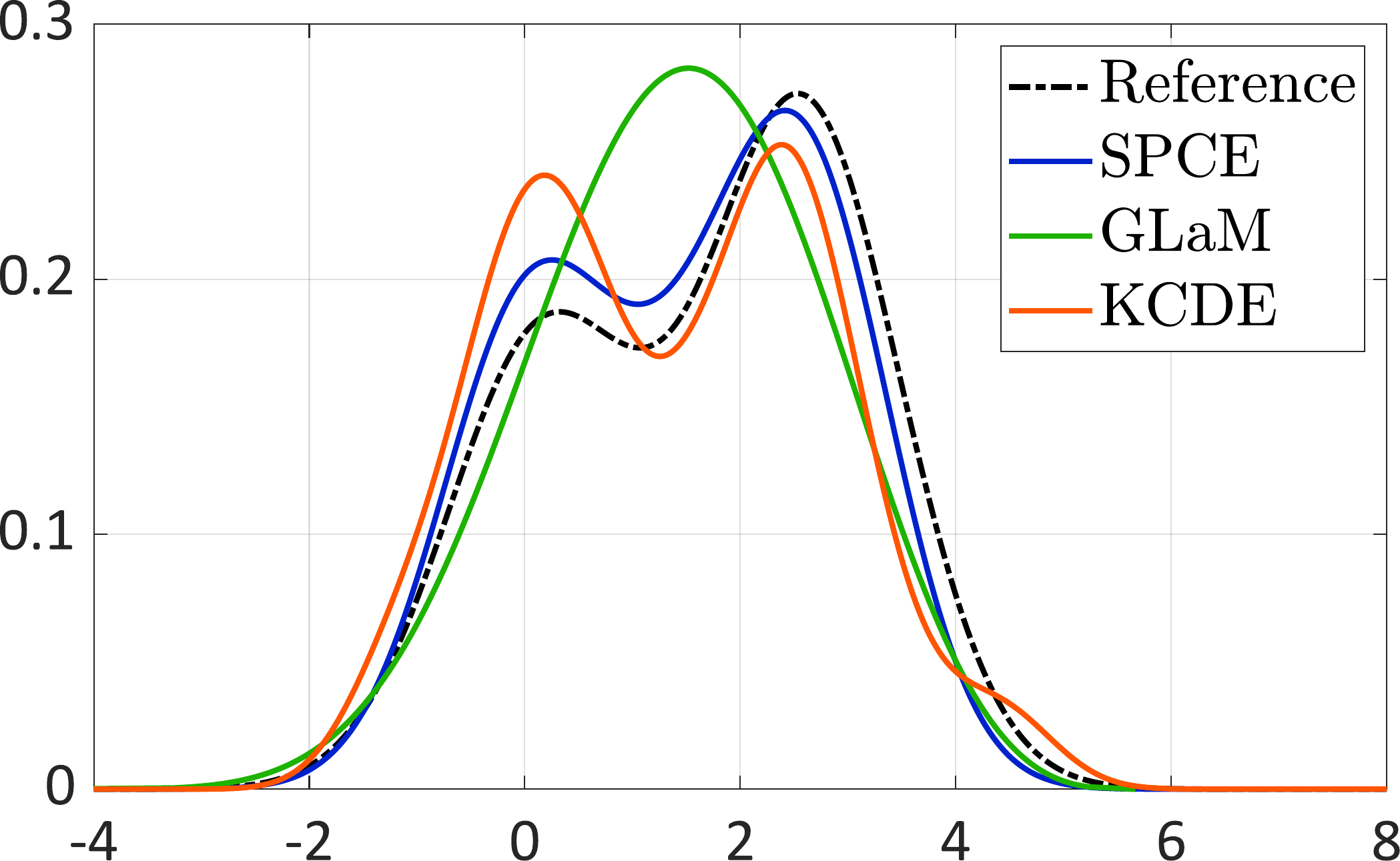}
		\caption{PDF for $x = 0.2$}
		\label{fig:bimodalpdf1}
	\end{subfigure}
	\begin{subfigure}[b]{.495\linewidth}
		\centering
		\includegraphics[height=0.58\linewidth, keepaspectratio]{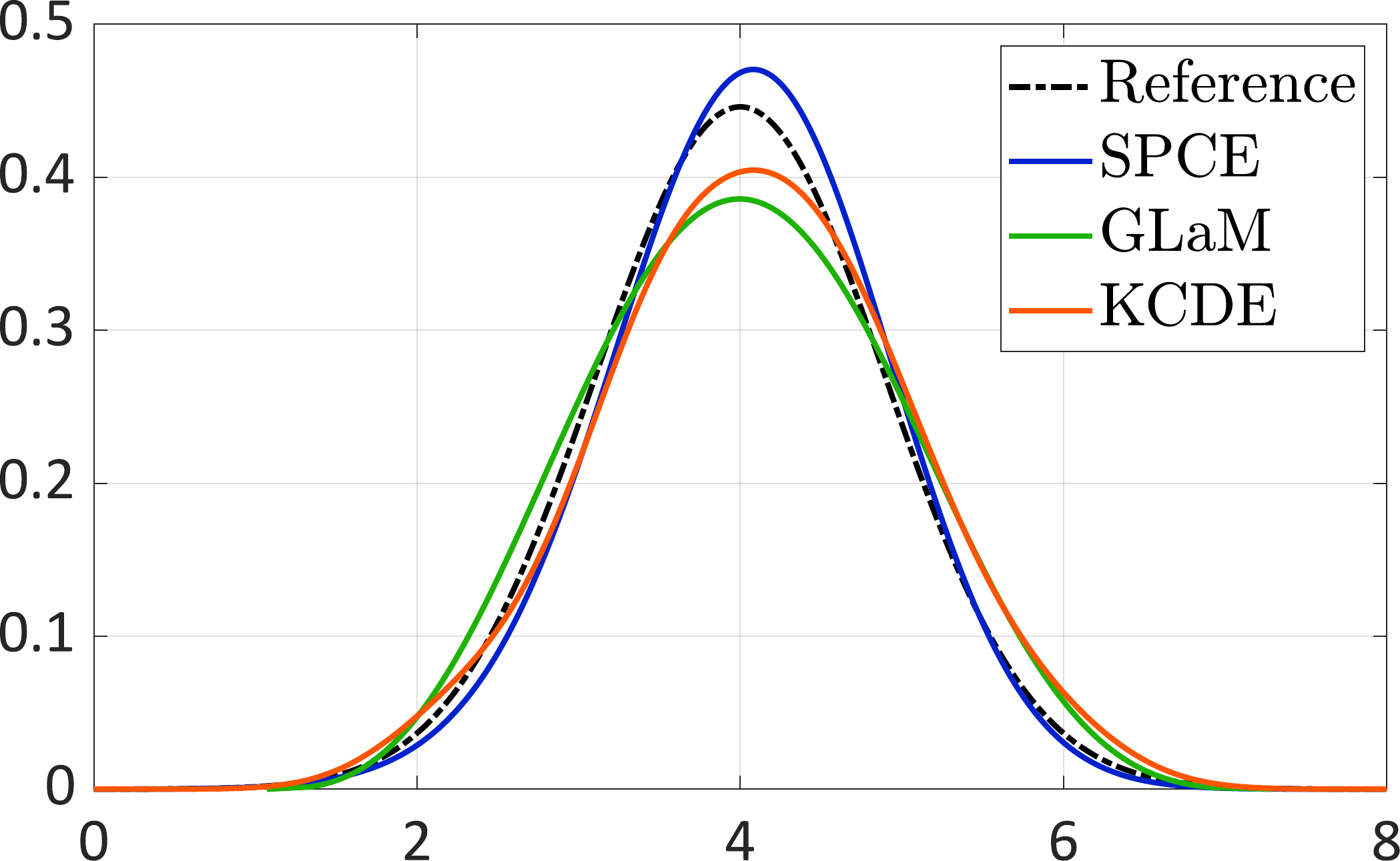}
		\caption{PDF for $x = 0.5$}
		\label{fig:bimodalpdf2}
	\end{subfigure}
	\begin{subfigure}[b]{.495\linewidth}
		\centering
		\includegraphics[height=0.58\linewidth, keepaspectratio]{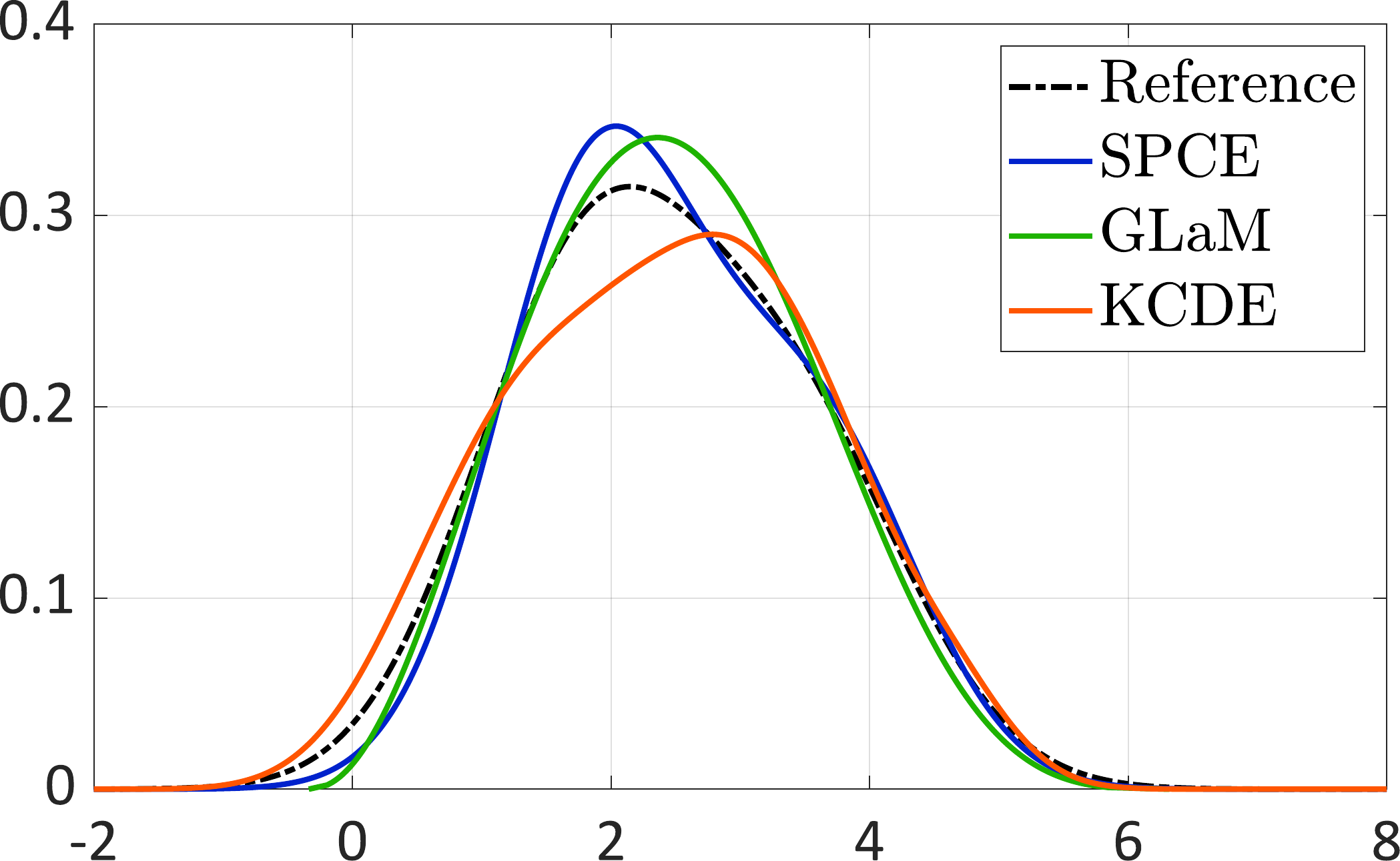}
		\caption{PDF for $x = 0.7$}
		\label{fig:bimodalpdf3}
	\end{subfigure}
	\begin{subfigure}[b]{.495\linewidth}
		\centering
		\includegraphics[height=0.58\linewidth, keepaspectratio]{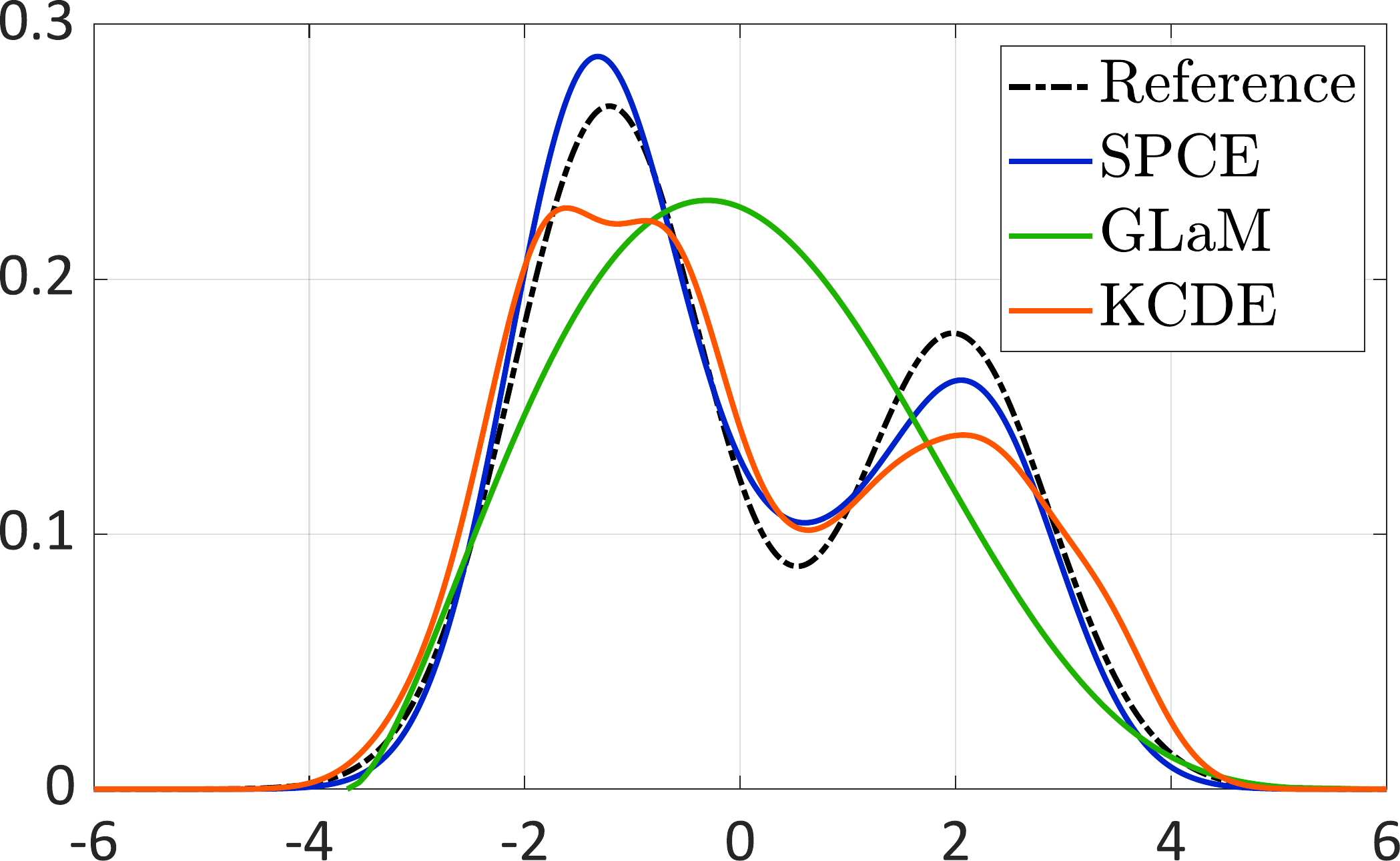}
		\caption{PDF for $x = 0.9$}
		\label{fig:bimodalpdf4}
	\end{subfigure}
	\caption{Bimodal analytical example --- Comparisons of the emulated PDFs, $N=800$.}
	\label{fig:bimodalpdf}
\end{figure}

\Cref{fig:bimodalpdf} compares the response PDF estimated by the surrogates built on the experimental design of \cref{fig:bimodaldata} ($N=800$) for four different values of $x$. We observe that small values of $x$ yield a bimodal distribution with the higher mode on the right. With $x$ increasing, the two modes merge and form a unimodal distribution at $x=0.5$. Then, the two modes separate again, which leads to bimodal distributions with the higher mode on the left. This shape variation can also be observed from \cref{fig:bimodaldata}. 

As opposed to the previous two examples, GLaM cannot represent this evolution, since generalized lambda distributions cannot produce multimodal distributions. In contrast, SPCE and KCDE capture well the bimodality and also the shape variation. Moreover, in \cref{fig:bimodalpdf3} the higher mode is moving to the left, which is a feature not exhibited by KCDE but correctly captured by SPCE.

\begin{figure}[!htbp]
	\centering
	\includegraphics[width=0.45\linewidth, keepaspectratio]{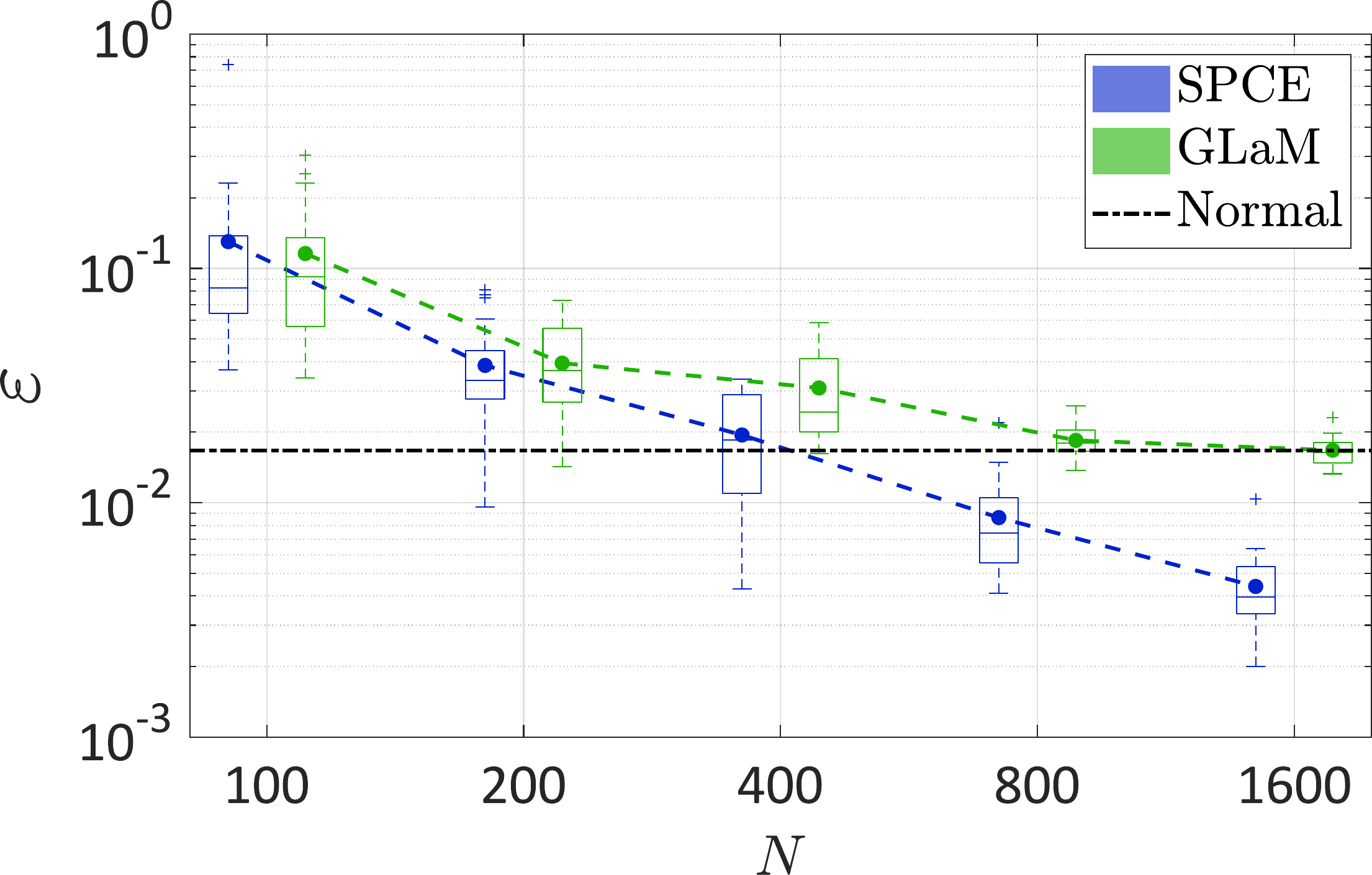}
	\hspace{5mm}
	\includegraphics[width=0.45\linewidth, keepaspectratio]{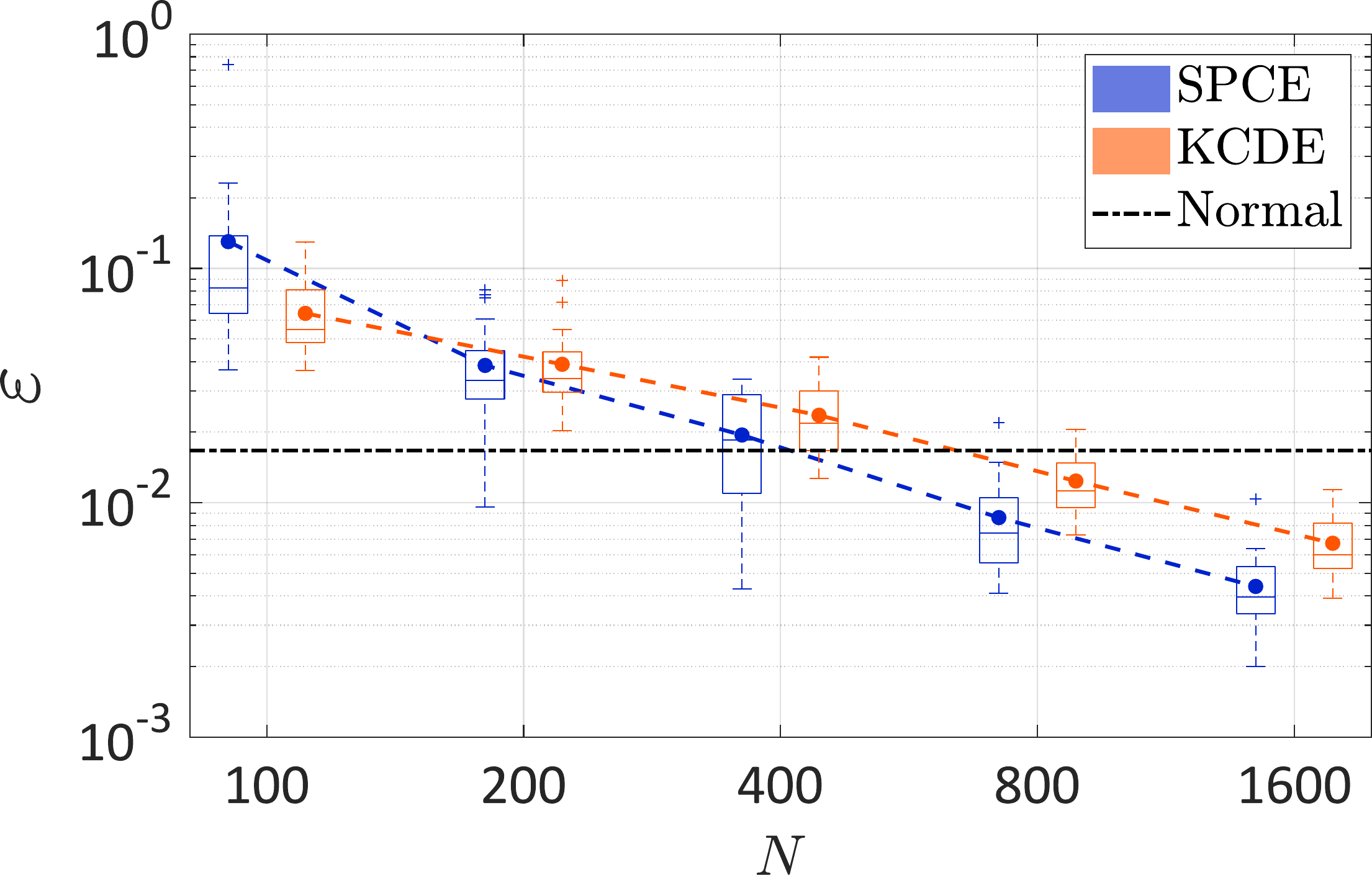}
	\caption{Bimodal analytical example --- Comparison of the convergence of the surrogate models. The dashed lines denote the average value over 20 repetitions of the full analysis. The black dash-dotted line represents the error of the model assuming that the response distribution is normal with the true mean and variance.}
	\label{fig:bimodal_WS}
\end{figure}

Quantitative comparisons for $N \in \acc{100;200;400;800;1{,}600}$ in \cref{fig:bimodal_WS} confirm our observation in \cref{fig:bimodalpdf}. Because of the bimodality, GLaM provides the least accurate approximation. When increasing $N$, it converges slowly to the same error as the best normal approximation which is clearly outperformed by the best two surrogates: SPCE and KCDE for $N\geq 800$. Both SPCE and KCDE show a consistent decay of the error. Only when a few samples $N=100$ are available does KCDE provide stabler estimates (the spread of the error is small) and better average performance. For $N \geq 200$, SPCE yields more accurate results and exhibits an overall faster rate of convergence. In summary, this example demonstrates that SPCE can represent bimodal distributions with a high accuracy.

\section{Conclusions}
\label{sec:conclusion}
In this paper, we present a novel surrogate model called stochastic polynomial chaos expansions (SPCE) to emulate the response distribution of stochastic simulators. This surrogate is an extension of the classical polynomial chaos expansions developed for deterministic simulators. In order to represent the intrinsic stochasticity of the simulator, we combine a latent variable with the well-defined inputs to form a polynomial chaos representation. In addition, we introduce an additive Gaussian noise as a regularizer. We propose using the maximum likelihood estimation for calibrating the coefficients $\ve{c}$ of the polynomial basis. The standard deviation $\sigma$ of the noise variable is a hyperparameter that regularizes the optimization problem for the polynomial coefficients $\ve{c}$ and is tuned by cross-validation to avoid overfitting. The cross-validation score is also used as a model selection criterion to choose an appropriate truncation scheme for the polynomial chaos expansion in an adaptive manner, and the most suitable distribution for the latent variable. As seen from the presentation and the application examples, the proposed method does not require replications. 

The performance of the developed method is illustrated on examples from mathematical finance and epidemiology and on an analytical example showcasing a bimodal response distribution. The results show that SPCE is able to well approximate various response distributions whether unimodal or not, with a reasonable number of model runs. 

Using an appropriate error measure defined in \cref{eq:Rlevel1}, SPCE is compared with the generalized lambda model (GLaM) and one state-of-the-art kernel conditional density estimator (KCDE). In the first two examples where the response distribution is unimodal, SPCE noticeably outperforms KCDE and provides slightly more accurate results than GLaM which is known for its flexibility for representing unimodal distributions. In the last example featuring bimodal distributions which cannot be well approximated by generalized lambda distributions, SPCE can still capture the complex shape variation and yields smaller errors than KCDE. All in all, SPCE generally performs as the best against the various competitors considered in this study.

Applications of the proposed method to complex engineering problems, such as wind turbine design \cite{AbdallahPEM2019} and structural dynamics \cite{MaiKonakli2017}, should be considered in future investigations. Statistical properties (e.g., consistency and asymptotics) of the maximum likelihood estimation used in SPCE remains to be studied. This will allow for assessing the uncertainty in the estimation procedure. 

Finally, the proposed approach has been validated so far only for problems with small to moderate dimentionality. To improve the efficiency and performance of SPCE in high dimensions, models that have a general sparse structure (not only regarding the mean function) are currently under investigations.


\section*{Acknowledgments}
This paper is a part of the project ``Surrogate Modeling for Stochastic 
Simulators (SAMOS)'' funded by the Swiss National Science Foundation (Grant 
\#200021\_175524), whose support is gratefully acknowledged.



\appendix
\section{Appendix}
\subsection{Upper bound}
\label{sec:upper}
In this section, we demonstrate that the leave-one-out error obtained from fitting the mean function \cref{eq:meanfit} provides an upper bound for $\sigma^2$.

Taking the expectation of \cref{eq:varYx} with respect to $\ve{X}$, it holds
\begin{equation}\label{eq:varY}
	\Esp{\Var{\tilde{Y} \con \ve{X}}} = \Esp{\sum_{\ve{\alpha} \in \caa\setminus\caa_m } c^2_{\ve{\alpha}} \psi^2_{\ve{\alpha}}(\ve{X}) + \sigma^2} = \sum_{\ve{\alpha} \in \caa\setminus\caa_m } c^2_{\ve{\alpha}} + \sigma^2.
\end{equation}

The leave-one-out error $\varepsilon_{\rm LOO}$ in the mean-fitting process is an estimate of $\Esp{\left(\hat{m}(\ve{X}) - Y_{\ve{X}}\right)^2}$ \cite{James2014}. The latter can be decomposed as
\begin{equation}
	\begin{split}
		\Esp{\left(\hat{m}(\ve{X}) - Y_{\ve{X}}\right)^2} &=\Esp{\left(\hat{m}(\ve{X}) - m(\ve{X}) + m(\ve{X}) - Y_{\ve{X}}\right)^2} \\
		&=\Esp{\left(\hat{m}(\ve{X})-m(\ve{X})\right)^2} + \Esp{\Var{Y\con \ve{X}}}
	\end{split}.
\end{equation}
Aiming at approximating $Y_{\ve{x}}$ with $\tilde{Y}_{\ve{x}}$, we have $\Esp{\Var{Y\con \ve{X}}} \approx \Esp{\Var{\tilde{Y} \con \ve{X}}}$. Hence, $\varepsilon_{\rm LOO}$ provides an upper bound for \cref{eq:varY} and therefore for $\sigma^2$.

\bibliography{References}

\begin{thebibliography}{10}

\bibitem{McNeil2005}
A.~J. McNeil, R.~Frey, and P.~Embrechts.
\newblock {\em Quantitative Risk Management: Concepts, Techniques, and Tools}.
\newblock Princeton Series in Finance. Princeton University Press, Princeton,
  New Jersey, 2005.

\bibitem{Britton2010}
T.~Britton.
\newblock Stochastic epidemic models: a survey.
\newblock {\em Math. Biosci.}, 225:24--35, 2010.

\bibitem{Ghanembook2003}
R.~Ghanem and P.~Spanos.
\newblock {\em Stochastic {F}inite {E}lements: {A} {S}pectral {A}pproach}.
\newblock {Courier} {D}over {P}ublications, Mineola, 2nd edition, 2003.

\bibitem{Rasmussen2006}
C.~E. Rasmussen and C.~K.~I. Williams.
\newblock {\em Gaussian processes for machine learning}.
\newblock Adaptive computation and machine learning. MIT Press, Cambridge,
  Massachusetts, {Internet} edition, 2006.

\bibitem{Ankenman2010}
B.~Ankenman, B.L. Nelson, and J.~Staum.
\newblock Stochastic {K}riging for simulation metamodeling.
\newblock {\em Oper. Res.}, 58:371--382, 2010.

\bibitem{Marrel2012}
A.~Marrel, B.~Iooss, S.~{Da~Veiga}, and M.~Ribatet.
\newblock Global sensitivity analysis of stochastic computer models with joint
  metamodels.
\newblock {\em Stat. Comput.}, 22:833--847, 2012.

\bibitem{Wooldridge2013}
J.~M. Wooldridge.
\newblock {\em Introductory Econometrics: A Modern Approach}.
\newblock Cengage Learning, 5th edition, 2013.

\bibitem{Binois2018}
M.~Binois, R.~B. Gramacy, and M.~Ludkovski.
\newblock Practical heteroscedastic {G}aussian process modeling for large
  simulation experiments.
\newblock {\em J. Comput. Graph. Stat.}, 27:808--821, 2018.

\bibitem{Koenker1978}
R.~Koenker and G.~Bassett.
\newblock Regression quantiles.
\newblock {\em Econometrica}, 46:33--50, 1978.

\bibitem{Plumlee2014}
M.~Plumlee and R.~Tuo.
\newblock Building accurate emulators for stochastic simulations via quantile
  {K}riging.
\newblock {\em Technometrics}, 56:466--473, 2014.

\bibitem{Torossian2020}
L.~Torossian, V.~Picheny, R.~Faivre, and A.~Garivier.
\newblock A review on quantile regression for stochastic computer experiments.
\newblock {\em Reliab. Eng. Sys. Safety}, 201, 2020.

\bibitem{Moutoussamy2015}
V.~Moutoussamy, S.~Nanty, and B.~Pauwels.
\newblock Emulators for stochastic simulation codes.
\newblock {\em ESAIM: Math. Model. Num. Anal.}, 48:116--155, 2015.

\bibitem{ZhuIJUQ2020}
X.~Zhu and B.~Sudret.
\newblock Replication-based emulation of the response distribution of
  stochastic simulators using generalized lambda distributions.
\newblock {\em Int. J. Uncertainty Quantification}, 10:249--275, 2020.

\bibitem{Browne2016}
T.~Browne, B.~Iooss, L.~{Le~Gratiet}, J.~Lonchampt, and E.~R\'emy.
\newblock Stochastic simulators based optimization by {G}aussian process
  metamodels -- application to maintenance investments planning issues.
\newblock {\em Quality Reliab. Eng. Int.}, 32(6):2067--2080, 2016.

\bibitem{Azzi2019}
S.~Azzi, B.~Sudret, and J.~Wiart.
\newblock Surrogate modeling of stochastic functions - application to
  computational electromagnetic dosimetry.
\newblock {\em Int. J. Uncertainty Quantification}, 9:351--363, 2019.

\bibitem{LuethenPREM2022}
N.~L\"uthen, S.~Marelli, and B.~Sudret.
\newblock Surrogates of stochastic simulators using trajectories.
\newblock {\em Prob. Eng. Mech.}, 2022.
\newblock (in preparation).

\bibitem{McCullagh1989}
P.~McCullagh and J.~Nelder.
\newblock {\em Generalized Linear Models}, volume~37 of {\em Monographs on
  Statistics and Applied Probability}.
\newblock Chapman and Hall/CRC, 1989.

\bibitem{Hastie1990}
T.~Hastie and R.~Tibshirani.
\newblock {\em Generalized Additive Models}, volume~43 of {\em Monographs on
  Statistics and Applied Probability}.
\newblock Chapman and Hall/CRC, 1990.

\bibitem{Fan1996}
J.~Fan and I.~Gijbels.
\newblock {\em Local Polynomial Modelling and Its Applications}, volume~66 of
  {\em Monographs on Statistics and Applied Probability}.
\newblock Chapman and Hall/CRC, 1996.

\bibitem{Hall2004}
P.~Hall, J.~Racine, and Q.~Li.
\newblock Cross-validation and the estimation of conditional probability
  densities.
\newblock {\em J. Amer. Stat. Assoc.}, 99:1015--1026, 2004.

\bibitem{Efromovich2010}
S.~Efromovich.
\newblock Dimension reduction and adaptation in conditional density estimation.
\newblock {\em J. Amer. Stat. Assoc.}, 105:761--774, 2010.

\bibitem{ZhuSIAMUQ2021}
X.~Zhu and B.~Sudret.
\newblock Emulation of stochastic simulators using generalized lambda models.
\newblock {\em SIAM/ASA J. Unc. Quant.}, 9:1345--1380, 2021.

\bibitem{ZhuRESS2021}
X.~Zhu and B.~Sudret.
\newblock Global sensitivity analysis for stochastic simulators based on
  generalized lambda surrogate models.
\newblock {\em Reliab. Eng. Sys. Safety}, 214(107815), 2021.

\bibitem{Nataf1962}
A.~Nataf.
\newblock D\'etermination des distributions dont les marges sont donn\'ees.
\newblock {\em C. R. Acad. Sci. Paris}, 225:42--43, 1962.

\bibitem{Rosenblatt1952}
M.~Rosenblatt.
\newblock Remarks on a multivariate transformation.
\newblock {\em Ann. Math. Stat.}, 23:470--472, 1952.

\bibitem{BlatmanPEM2010}
G.~Blatman and B.~Sudret.
\newblock An adaptive algorithm to build up sparse polynomial chaos expansions
  for stochastic finite element analysis.
\newblock {\em Prob. Eng. Mech.}, 25:183--197, 2010.

\bibitem{Ernst2012}
O.~G. Ernst, A.~Mugler, H.~J. Starkloff, and E.~Ullmann.
\newblock On the convergence of generalized polynomial chaos expansions.
\newblock {\em ESAIM: Math. Model. and Num. Anal.}, 46:317--339, 2012.

\bibitem{Xiu2002}
D.~Xiu and G.~E. Karniadakis.
\newblock {The Wiener-Askey polynomial chaos for stochastic differential
  equations}.
\newblock {\em SIAM J. Sci. Comput.}, 24(2):619--644, 2002.

\bibitem{Gautschi2004}
W.~Gautschi.
\newblock {\em Orthogonal polynomials: computation and approximation}.
\newblock Oxford University Press, 2004.

\bibitem{SudretJCP2011}
G.~Blatman and B.~Sudret.
\newblock Adaptive sparse polynomial chaos expansion based on {L}east {A}ngle
  {R}egression.
\newblock {\em J. Comput. Phys.}, 230:2345--2367, 2011.

\bibitem{Berveiller2006}
M.~Berveiller, B.~Sudret, and M.~Lemaire.
\newblock Stochastic finite elements: a non intrusive approach by regression.
\newblock {\em Eur. J. Comput. Mech.}, 15(1-3):81--92, 2006.

\bibitem{Doostan2011}
A.~Doostan and H.~Owhadi.
\newblock A non-adapted sparse approximation of {PDEs} with stochastic inputs.
\newblock {\em J. Comput. Phys.}, 230(8):3015--3034, 2011.

\bibitem{Babacan2010}
S.D. Babacan, R.~Molina, and A.K. Katsaggelos.
\newblock Bayesian compressive sensing using {L}aplace priors.
\newblock {\em IEEE Trans. Image Process.}, 19(1):53--63, 2010.

\bibitem{Luethen2021}
N.~L\"uthen, S.~Marelli, and B.~Sudret.
\newblock Sparse polynomial chaos expansions: {Literature} survey and
  benchmark.
\newblock {\em SIAM/ASA J. Unc. Quant.}, 9(2):593--649, 2021.

\bibitem{LuethenIJUQ2021}
N.~L\"uthen, S.~Marelli, and B.~Sudret.
\newblock A benchmark of basis-adaptive sparse polynomial chaos expansions for
  engineering regression problems.
\newblock {\em Int. J. Uncertainty Quantification}, 2021.
\newblock (submitted).

\bibitem{SudretRESS2008b}
B.~Sudret.
\newblock Global sensitivity analysis using polynomial chaos expansions.
\newblock {\em Reliab. Eng. Sys. Safety}, 93:964--979, 2008.

\bibitem{Everitt1984}
B.~S. Everitt.
\newblock {\em An Introduction to Latent Variables Models}.
\newblock Chapman \& Hall, 1984.

\bibitem{Desceliers2006a}
C.~Desceliers, R.~Ghanem, and C.~Soize.
\newblock Maximum likelihood estimation of stochastic chaos representations
  from experimental data.
\newblock {\em Int. J. Numer. Meth. Engng.}, 66:978--1001, 2006.

\bibitem{Jacod2004}
J.~Jacod and P.~Protter.
\newblock {\em Probability Essentials}.
\newblock Springer, 2nd edition, 2004.

\bibitem{Hastie2001}
T.~Hastie, R.~Tibshirani, and J.~Friedman.
\newblock {\em The elements of statistical learning: {D}ata mining, inference
  and prediction}.
\newblock Springer, New York, 2001.

\bibitem{Golub1969}
G.~H. Golub and J.~H. Welsch.
\newblock Calculation of {G}auss quadrature rules.
\newblock {\em Mathematics of computation}, 23(106):221--230, 1969.

\bibitem{Fletcher1987}
R.~Fletcher.
\newblock {\em Practical Methods of Optimization}.
\newblock John Wiley \& Sons, 2nd edition, 1987.

\bibitem{Snoek2012}
J.~Snoek, H.~Larochelle, and R.~P. Adams.
\newblock Practical {B}ayesian optimization of machine learning algorithms.
\newblock In F.~Pereira, C.~J.~C. Burges, L.~Bottou, and K.~Q. Weinberger,
  editors, {\em Advances in Neural Information Processing Systems 25}, pages
  2951--2959. Curran Associates, Inc., 2012.

\bibitem{Hayfield2008}
T.~Hayfield and J.S. Racine.
\newblock {Nonparametric Econometrics: The np Package}.
\newblock {\em J. Stat. Softw.}, 2008.

\bibitem{Villani2008}
C.~Villani.
\newblock {\em Optimal transport, old and new}.
\newblock Cambridge Series in Statistical and Probabilistic Mathematics.
  Springer, Cambridge, 2000.

\bibitem{McKay1979}
M.~D. McKay, R.~J. Beckman, and W.~J. Conover.
\newblock A comparison of three methods for selecting values of input variables
  in the analysis of output from a computer code.
\newblock {\em Technometrics}, 21(2):239--245, 1979.

\bibitem{Reddy2016}
K.~Reddy and V.~Clinton.
\newblock Simulating stock prices using geometric {B}rownian motion: Evidence
  from {A}ustralian companies.
\newblock {\em Australasian Accounting, Business and Finance Journal},
  10(3):23--47, 2016.

\bibitem{Shreve2004}
S.~Shreve.
\newblock {\em Stochastic Calculus for Finance {II}}.
\newblock Springer, New York, 2004.

\bibitem{Gillespie1977}
D.~T. Gillespie.
\newblock Exact stochastic simulation of coupled chemical reactions.
\newblock {\em J. Phys. Chem.}, 81:2340--2361, 1977.

\bibitem{AbdallahPEM2019}
I.~Abdallah, C.~Lataniotis, and B.~Sudret.
\newblock Parametric hierarchical {K}riging for multi-fidelity
  aero-servo-elastic simulators -- application to extreme loads on wind
  turbines.
\newblock {\em Prob. Eng. Mech.}, 55:67--77, 2019.

\bibitem{MaiKonakli2017}
C.~V. Mai, K.~Konakli, and B.~Sudret.
\newblock Seismic fragility curves for structures using non-parametric
  representations.
\newblock {\em Frontiers Struct. Civ. Eng.}, 11(2), 2017.

\bibitem{James2014}
G.~James, D.~Witten, T.~Hastie, and R.~Tibshirani.
\newblock {\em An Introduction to Statistical Learning: with Applications in
  {R}}.
\newblock Springer, 2014.

\end{thebibliography}

\end{document}